\documentclass[onecolumn]{aastex631}

\newcommand\codename{\texttt{GWFAST}}
\newcommand\wfname{\texttt{WF4Py}}
\newcommand{\codestring}[1]{\textcolor{red!80!black}{\texttt{\textquotesingle #1\textquotesingle}}}
\newcommand{\codekeyword}[1]{\textcolor{green!50!black}{\texttt{#1}}}
\usepackage[english]{babel}
\usepackage[utf8]{inputenc}
\usepackage{savesym}
\usepackage{comment}
\usepackage{nccmath}
\usepackage{bbold, amsmath}
\usepackage[caption=false]{subfig}
\usepackage{enumitem}
\usepackage{xcolor}
\definecolor{neonpurple}{RGB}{176, 38, 255}
\definecolor{lightblue}{RGB}{114, 189, 212}
\definecolor{brightgreen}{RGB}{72, 179, 0}
\usepackage{physics}
\usepackage{nicefrac}
\usepackage{mathtools}
\usepackage{booktabs}
\usepackage{footmisc}
\usepackage{multirow}
\savesymbol{tablenum}
\usepackage{siunitx}
\restoresymbol{SIX}{tablenum}
\DeclareSIUnit \parsec {pc}
\DeclareSIUnit \arcsecondfull {arcsec}
\DeclareSIUnit \year{yr}
\DeclareSIUnit \day{day}
\DeclareSIUnit \hour{hr}
\DeclareSIUnit \radiant{rad}
\DeclareSIUnit \degfull{deg}
\DeclareSIUnit \erg {erg}
\DeclareSIUnit \Lsun {L_\odot}
\DeclareSIUnit \Msun {M_\odot}
\DeclareSIUnit \AstroUnit {au}
\usepackage{letltxmacro}
\LetLtxMacro{\originaleqref}{\eqref}
\renewcommand{\eqref}{Eq.~\originaleqref}

\addto\extrasenglish{%
  \renewcommand{\footnoteautorefname}{footnote}%
}

\LetLtxMacro{\originalfootref}{\footref}
\renewcommand{\footref}{\footnoteautorefname~\ref}
\usepackage{listings}
\lstdefinestyle{mystyle}{
    backgroundcolor=\color{gray!3!white},   
    commentstyle=\color{lightblue},
    keywordstyle=\color{green!50!black},
    numberstyle=\tiny\color{black},
    stringstyle=\color{red!80!black},
    basicstyle=\ttfamily\small,
    breakatwhitespace=true,         
    breaklines=true,                 
    captionpos=b,                    
    keepspaces=false,                
    numbers=none,                    
    numbersep=5pt,                  
    showspaces=false,                
    showstringspaces=false,
    showtabs=false,                  
    tabsize=2,
    frame=lines,
    upquote=true,
    language=Python,
}
\newcommand\digitstyle{\color{brightgreen}}
\makeatletter
\newcommand{\ProcessDigit}[1]
{%
  \ifnum\lst@mode=\lst@Pmode\relax%
   {\digitstyle #1}%
  \else
    #1%
  \fi
}
\makeatother
\def\stopsqrt{\mathpalette\DHLhksqrt}
\def\DHLhksqrt#1#2{\setbox0=\hbox{$#1\sqrt{#2\,}$}\dimen0=\ht0
	\advance\dimen0-0.2\ht0
	\setbox2=\hbox{\vrule height\ht0 depth -\dimen0}%
	{\box0\lower0.4pt\box2}}

\shorttitle{\codename{}: a Fisher information matrix \texttt{Python} code for third--generation gravitational--wave detectors}
\shortauthors{Iacovelli, Mancarella, Foffa, Maggiore}

\begin{document}

\reportnum{ET-0142A-22}

\title{\codename{}: a Fisher information matrix \texttt{Python} code for third--generation gravitational--wave detectors}

\correspondingauthor{Francesco Iacovelli}
\email{Francesco.Iacovelli@unige.ch}

\author[0000-0002-4875-5862]{Francesco Iacovelli}
\affiliation{D\'epartement de Physique Th\'eorique, Universit\'e de Gen\`eve, 24~quai Ernest Ansermet, 1211~Gen\`eve~4, Switzerland}

\author[0000-0002-0675-508X]{Michele Mancarella}
\affiliation{D\'epartement de Physique Th\'eorique, Universit\'e de Gen\`eve, 24~quai Ernest Ansermet, 1211~Gen\`eve~4, Switzerland}

\author[0000-0002-4530-3051]{Stefano Foffa}
\affiliation{D\'epartement de Physique Th\'eorique, Universit\'e de Gen\`eve, 24~quai Ernest Ansermet, 1211~Gen\`eve~4, Switzerland}

\author[0000-0001-7348-047X]{Michele Maggiore}
\affiliation{D\'epartement de Physique Th\'eorique, Universit\'e de Gen\`eve, 24~quai Ernest Ansermet, 1211~Gen\`eve~4, Switzerland}

\begin{abstract}

We introduce \codename{}~\raisebox{-1pt}{\href{https://github.com/CosmoStatGW/gwfast}{\includegraphics[width=10pt]{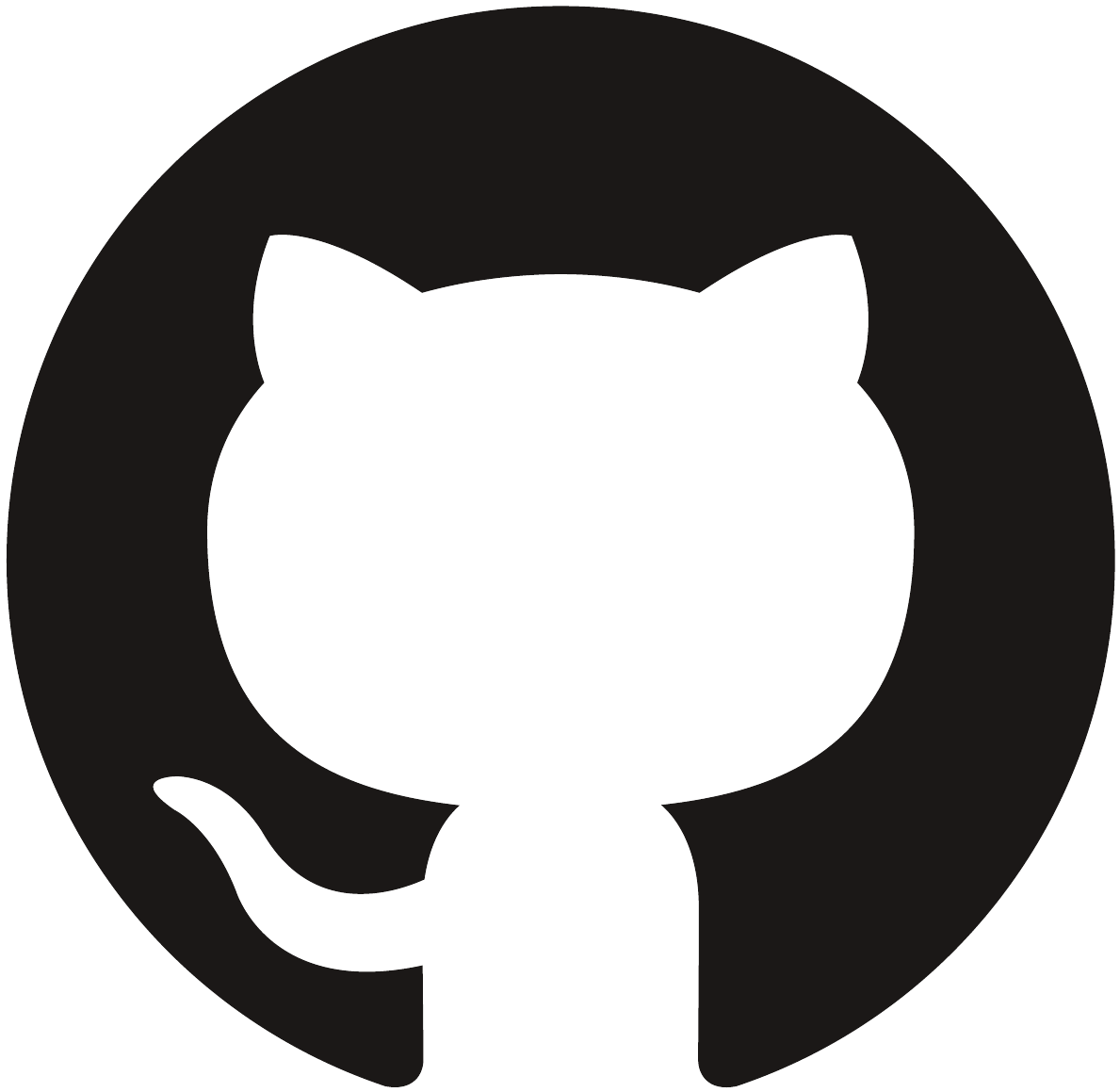}}}~\footnote{\url{https://github.com/CosmoStatGW/gwfast}}, a Fisher information matrix \texttt{Python} code that allows easy and efficient estimation of signal--to--noise ratios and parameter measurement errors for large catalogs of resolved sources observed by networks of gravitational–wave detectors. 
In particular, \codename{} includes the effects of the Earth's motion during the evolution of the signal, supports parallel computation, and relies on automatic differentiation rather than on finite differences techniques, which allows the computation of derivatives with accuracy close  to machine precision.
We also release the library \wfname{}~\raisebox{-1pt}{\href{https://github.com/CosmoStatGW/WF4Py}{\includegraphics[width=10pt]{GitHub-Mark.pdf}}}~\footnote{\url{https://github.com/CosmoStatGW/WF4Py}} implementing state--of--the--art gravitational--wave waveforms in \texttt{Python}. In this paper we provide a documentation of \codename{} and \wfname{} with practical examples and tests of performance and reliability. In the companion paper \cite{Iacovelli:2022bbs} we present forecasts for the detection capabilities of the second and third generation of ground--based gravitational--wave detectors, obtained with \codename{}.

\end{abstract}

\section{Outline}

\begin{figure}[t]
    \centering
    \includegraphics[width=1.\textwidth]{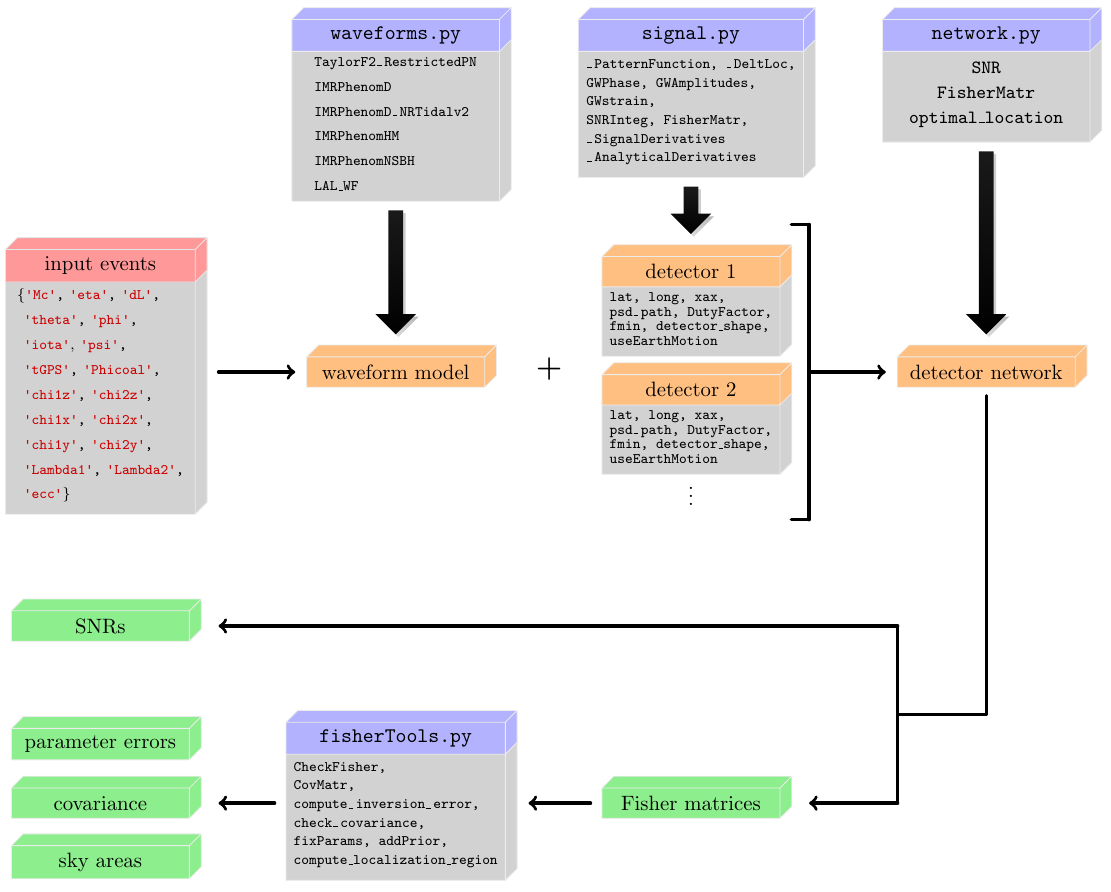}
    \caption{Flowchart of the functioning of \codename{}. See \autoref{sec:input} for the description of the inputs.}
    \label{fig:GWFast_Scheme}
\end{figure}

\codename{} is a new, fast and accurate software, capable of computing signal--to--noise ratios (SNRs) and parameter measurement errors for networks of gravitational--wave (GW) detectors, using the Fisher Information Matrix (FIM) formalism. This approximates the full posterior probability distribution for the parameters of a GW signal (see e.g. \cite{Cutler:1994ys,Vallisneri:2007ev,Rodriguez:2013mla} for a comprehensive treatment) and is used for forecasts on large catalogs of sources for which a full parameter estimation would be computationally too expensive. The computational cost is the main limitation of present--day forecast studies, especially for the third generation of GW detectors. 
This is related to two main aspects. The first is the duration of the signal (in particular, for binary neutron stars at ground--based detectors), which requires to correctly account for the time evolution of the antenna pattern functions and makes the data analysis challenging in terms of computational resources. To our knowledge, the problem of a full Bayesian inference for even a single one of such events is not manageable with techniques and resources used for second--generation (2G) detectors. Only recently dedicated approaches have started to be investigated \citep{Smith:2021bqc}. The second aspect is the scalability to large catalogs. The study of the reach and parameter estimation capabilities of third--generation (3G) detectors is a key aspect for assessing their scientific potential, and typically requires to study catalogs of tens of thousands of sources. 
\codename{} is suitable for these applications since it accounts for state--of--the--art waveform models, the effect of the motion of the Earth, and the possibility of parallel evaluations when running on large catalogs. Moreover, it does not rely on finite difference techniques to compute derivatives, but on automatic differentiation, which is a method that does not suffer from possible inaccuracies arising from the computation of the derivatives, in particular related to the choice of the step size. Hence we make it publicly available, together with routines to run in parallel. In this paper we provide a documentation, tests to validate the reliability of the code, and some examples.
A scheme of the organization of the code is reported in \autoref{fig:GWFast_Scheme}.
In the companion paper \cite{Iacovelli:2022bbs} we used \codename{} to produce forecasts for the detection capabilities  of LIGO--Virgo--KAGRA (LVK) during their forthcoming O4 run, and of 3G ground--based GW detectors, namely, Einstein Telescope (ET) and Cosmic Explorer (CE), based on up--to--date models of the expected population of sources. 

This paper is structured as follows. In \autoref{sec:signalModel} we describe the conventions for the input parameters and the waveform models available in \codename{}, which are a pure \texttt{Python} version of those contained in the LIGO Algorithm Library \texttt{LAL} \citep{lalsuite}, and compare with their original implementation. The waveform models are also separately released in a ready--to--use version, \wfname{}. Moreover, \codename{} implements an interface with \texttt{LAL}, so that all the waveforms available in this library can be used directly. Only the use of the pure \texttt{Python} implementation allows however to fully exploit the vectorization capabilities of \codename{} and the use of automatic differentiation, as explained below. In \autoref{sec:sig_net} we document the two core modules of the software: \texttt{signal} and \texttt{network}, which allow the user to easily compute SNRs and FIMs, with various code examples. In particular, in \autoref{sec:signal_derivs} we provide an overview of how \codename{} deals with the computation of the derivatives of the signal with respect to the parameters. If using the  \texttt{Python} implementation of the waveforms, we evaluate those using the automatic differentiation module of the \texttt{JAX} library \citep{jax2018github}, which ensures a fast and accurate computation, while if using the \texttt{LAL} waveforms the computation is performed using finite differences techniques. In \autoref{sec:CovMatr} we then describe how \codename{} deals with the inversion of the Fisher matrix, to obtain the covariance matrix and thus the measurement errors, and the various manipulations that can be performed using the module \texttt{fisherTools}. \autoref{sec:parallel_run} is devoted to the description of how to run \codename{} on multiple CPUs, so to easily handle huge catalogs of events, through the module \texttt{calculate\_forecasts\_from\_catalog.py}. Finally in \autoref{sec:comparison_realEv}, to assess its reliability, we show the application of \codename{} to some of the real events observed by the LIGO and Virgo interferometers during their second and third observing runs, for which a full Bayesian parameter estimation has been performed. In \autoref{sec:summary} we then summarise and conclude.
\newpage
\section{Signal modelling}\label{sec:signalModel}
Waveform models give the prediction for a GW signal emitted by a coalescing binary as a function of the parameters of the source. We here give a brief overview of the chosen conventions for these parameters, as well as the waveforms available in \codename{} and their implementation.

\subsection{Input parameters}\label{sec:input}
In the most general case, restricting to quasi--circular orbits, the parameters needed to describe the signal emitted by a coalescing binary system are 15 for a binary black hole (BBH), 17 for a binary neutron star (BNS) and 16 for a neutron star black hole binary (NSBH), i.e. $\vb*{\theta} = \{{\cal M}_c, \eta, d_L, \theta, \phi, \iota, \psi, t_c, \Phi_c, \chi_{1,x}, \chi_{2,x}, \chi_{1,y}, \chi_{2,y}, \chi_{1,z}, \chi_{2,z}, \Lambda_1, \Lambda_2\}$ (see e.g. \cite{Maggiore:2007ulw}), where: ${\cal M}_c$ denotes the detector--frame chirp mass, $\eta$ the symmetric mass ratio, $d_L$ the luminosity distance to the source, $\theta$ and $\phi$ are the sky position coordinates, defined as $\theta=\pi/2-\delta$ and $\phi=\alpha$ (with $\alpha$ and $\delta$ right ascension and declination, respectively), $\iota$ the inclination angle of the binary orbital angular momentum with respect to the line of sight, $\psi$ the polarisation angle, $t_c$ the time of coalescence, $\Phi_c$ the phase at coalescence, $\chi_{i,c}$ the dimensionless spin of the object $i=\{1,2\}$ along the axis $c = \{x,y,z\}$ and $\Lambda_i$ the dimensionless tidal deformability of the object $i$, which is zero for a BH. \codename{} expresses the chirp mass in units of solar masses, \si{\Msun}, the luminosity distance in \si{\giga\parsec}, the time of coalescence as a GPS time, in seconds, and all the angles in radiants. Including also a small eccentricity in the orbit, there is one more parameter to consider, $e_0$, which denotes the eccentricity at a given reference frequency.\footnote{The reference frequency is chosen when initialising the waveform model, through the argument \texttt{fRef\_ecc}. The default is \texttt{fRef\_ecc\textcolor{neonpurple}{=}\codekeyword{None}}, meaning the minimum frequency of the frequency grid is used.\label{footnote:fRef_ecc}}

To use \codename{}, the parameters $\vb*{\theta}$ of the events in the catalog to analyse have to be stored in a dictionary, with the same keys as in the following example
\begin{lstlisting}[caption={Input parameters for waveform models and FIM calculations in \codename{}.},label={lst:events_dict_init}]
import numpy as np

events = {'Mc':np.array([...]), 'eta':np.array([...]), 'dL':np.array([...]),
          'theta':np.array([...]), 'phi':np.array([...]), 'iota':np.array([...]),
          'psi':np.array([...]), 'tGPS':np.array([...]), 'Phicoal':np.array([...]),
          'chi1z':np.array([...]), 'chi2z':np.array([...]),
          'chi1x':np.array([...]), 'chi2x':np.array([...]),
          'chi1y':np.array([...]), 'chi2y':np.array([...]),
          'Lambda1':np.array([...]), 'Lambda2':np.array([...]), 'ecc':np.array([...])
         }
\end{lstlisting}
It is also possible to pass some entries with different commonly used parametrizations, namely:
\begin{enumerate}[label=(\alph*)]
    \item\label{item:inp_radec} the sky position coordinates can be given in terms of right ascension $\alpha$ and declination $\delta$, always in radiants, in place of $\theta$ and $\phi$;
    \item\label{item:inp_tc} the time of coalescence can be provided as a Greenwich Mean Sidereal Time (GMST) in days, under the entry name \codestring{tcoal}, which takes the place of \codestring{tGPS};
    \item\label{item:inp_chiSA} in the non precessing case, one can choose the spin parameters $\chi_s, \chi_a$ instead of $\chi_{1,z}, \chi_{1,z}$, defined as
    \begin{equation}\label{eq:chiSchiA_def}
        \chi_s = \dfrac{1}{2}(\chi_{1,z} + \chi_{2,z})\,, \qquad \chi_a = \dfrac{1}{2}(\chi_{1,z} - \chi_{2,z}) \, ;
    \end{equation}
    \item\label{item:inp_LamT} it is possible to use the combinations of the tidal deformabilities $\tilde{\Lambda}, \delta\tilde{\Lambda}$ in place of $\Lambda_1, \Lambda_2$, whose definitions are \citep{PhysRevD.89.103012}
    \begin{subequations}\label{eq:LamTdelLam_def}
    \begin{align}
        \tilde{\Lambda} &= \dfrac{8}{13} \left[(1+7\eta-31\eta^2)(\Lambda_1 + \Lambda_2) + \sqrt{1-4\eta}(1+9\eta-11\eta^2)(\Lambda_1 - \Lambda_2)\right]\,,\\
        \delta\tilde{\Lambda} &= \dfrac{1}{2} \left[\sqrt{1-4\eta} \left(1-\dfrac{13272}{1319}\eta + \dfrac{8944}{1319}\eta^2\right)(\Lambda_1 + \Lambda_2) + \left(1 - \dfrac{15910}{1319}\eta + \dfrac{32850}{1319}\eta^2 + \dfrac{3380}{1319}\eta^3\right)(\Lambda_1 - \Lambda_2)\right]\,;
    \end{align}
    \end{subequations}
    \item\label{item:inp_prec} if using a waveform model which includes the contribution of unaligned spin components (precessing spins) it is possible to substitute the entries $\iota, \chi_{1,x}, \chi_{2,x}, \chi_{1,y}, \chi_{2,y}, \chi_{1,z}, \chi_{2,z}$ with $\theta_{JN}, \chi_{1}, \chi_{2}, \theta_{s, 1}, \theta_{s, 2}, \phi_{JL}, \phi_{1,2}$.
   These are, respectively, the angle between the total angular momentum and the line of sight, $\theta_{JN}$, the magnitudes of the spin vectors, $\chi_i$, the angles between the spin vectors and the orbital angular momentum, $\theta_{s,i}$, the azimuthal angle of the orbital angular momentum and the total angular momentum, $\phi_{JL}$, and the difference in azimuthal angle between the two spin vectors, $\phi_{1,2}$.
\end{enumerate}
A summary of the parameters, their physical symbol and their name in \codename{} is provided in \autoref{tab:input_pars}.

\begin{table}[b]
    \centering
    \begin{tabular}{!{\vrule width .09em}c|c|c|c|c!{\vrule width .09em}}
    \toprule\midrule
        parameter symbol & parameter description & name in \codename{} & units in \codename{} & physical range\\
        \midrule\midrule
         ${\cal M}_c$ & detector--frame chirp mass & \codestring{Mc} & \si{\Msun} & $(0,\, +\infty)$\\
         \midrule
         $\eta$ & symmetric mass ratio & \codestring{eta} & -- & $(0,\, 0.25]$\\
         \midrule
         $d_L$ & luminosity distance & \codestring{dL} & \si{\giga\parsec} & $(0,\, +\infty)$\\
         \midrule
         \multirow{2}{*}{$\theta$, $\phi$ / $\alpha$, $\delta \, ^{\rm \ref{item:inp_radec}}$} & \multirow{2}{*}{sky position} & \codestring{theta}, \codestring{phi} / & \multirow{2}{*}{rad} & $[0,\, \pi]$, $[0,\, 2\pi]$ / \\
         & & \codestring{ra}, \codestring{dec} & & $[0,\, 2 \pi]$, $[-\pi/2,\, \pi/2]$\\
         \midrule
         \multirow{2}{*}{$\iota\,^{\rm \ref{item:inp_prec}}$}& inclination angle w.r.t. orbital & \multirow{2}{*}{\codestring{iota}} & \multirow{2}{*}{rad} & \multirow{2}{*}{$[0,\, \pi]$}\\
        & angular momentum & & & \\
        \midrule
        $\psi$ & polarisation angle & \codestring{psi} & rad & $[0,\, \pi]$ \\
         \midrule
        $t_c\, ^{\rm \ref{item:inp_tc}}$ & time of coalescence GPS / GMST & \codestring{tGPS} / \codestring{tcoal} & s / day & $[0,\, +\infty)$ / $[0,\, 1]$\\
        \midrule
        $\Phi_c$ & phase at coalescence & \codestring{Phicoal} & rad & $[0,\, 2\pi]$\\
        \midrule
        \multirow{2}{*}{$\chi_{i,c} \, ^{\rm \ref{item:inp_chiSA},\, \ref{item:inp_prec}}$} & spin component of object $i=\{1,2\}$ & \codestring{chi1x}, \codestring{chi1y}, \codestring{chi1z} & \multirow{2}{*}{--} & \multirow{2}{*}{$[-1,\,1]$, ($S^2$)} \\
        & along axis $c=\{x,y,z\}$ & \codestring{chi2x}, \codestring{chi2y}, \codestring{chi2z} & & \\
        \midrule
        \multirow{2}{*}{$\Lambda_{i} \, ^{\rm \ref{item:inp_LamT}}$} & adimensional tidal deformability & \multirow{2}{*}{\codestring{Lambda1}, \codestring{Lambda2}} & \multirow{2}{*}{--} & \multirow{2}{*}{$[0,\, +\infty)$}\\
        & of object $i=\{1,2\}$ & & & \\
        \midrule
        $e_0$ & orbital eccentricity & \codestring{ecc} & --  & $[0,\, 1)$\\
        \midrule
        \multirow{2}{*}{$\chi_s$, $\chi_a \, ^{\rm \ref{item:inp_chiSA}}$} & symmetric and asymmetric spin & \multirow{2}{*}{\codestring{chiS}, \codestring{chiA}} & \multirow{2}{*}{--} & \multirow{2}{*}{$[-1,\, 1]$, $[-1,\, 1]$}\\
        & components, see \eqref{eq:chiSchiA_def} & & & \\
        \midrule
        \multirow{2}{*}{$\tilde{\Lambda}$, $\delta\tilde{\Lambda} \,^{\rm \ref{item:inp_LamT}}$} & adimensional tidal deformability & \codestring{LambdaTilde}, & \multirow{2}{*}{--} & $[0,\,+\infty)$,\\
        & combinations, see \eqref{eq:LamTdelLam_def} & \codestring{deltaLambda} & & $(-\infty,\,+\infty)$\\
        \midrule
        \multirow{2}{*}{$\theta_{JN}\, ^{\rm \ref{item:inp_prec}}$}& inclination angle w.r.t. total & \multirow{2}{*}{\codestring{thetaJN}} & \multirow{2}{*}{rad} & \multirow{2}{*}{$[0, \, \pi]$}\\
        & angular momentum & & & \\
        \midrule
        $\chi_i \, ^{\rm \ref{item:inp_prec}}$ & spin magnitude of object $i=\{1,2\}$ & \codestring{chi1}, \codestring{chi2} & -- & $[0,\, 1]$ \\
        \midrule
        $\theta_{s,i} \, ^{\rm \ref{item:inp_prec}}$ & spin tilt of object $i=\{1,2\}$ & \codestring{tilt1}, \codestring{tilt2} & rad & $[0,\, \pi]$\\
        \midrule
        \multirow{2}{*}{$\phi_{JL} \, ^{\rm \ref{item:inp_prec}}$} & azimuthal angle between orbital & \multirow{2}{*}{\codestring{phiJL}} & \multirow{2}{*}{rad} & \multirow{2}{*}{$[0,\, 2\pi]$}\\
        & and total angular momentum & & & \\
        \midrule
        \multirow{2}{*}{$\phi_{1,2} \, ^{\rm \ref{item:inp_prec}}$} & difference in azimuthal angle & \multirow{2}{*}{\codestring{phi12}} & \multirow{2}{*}{rad} & \multirow{2}{*}{$[0,\, 2\pi]$}\\
        & between the spin vectors & & & \\
    \midrule\bottomrule
    \end{tabular}
    \caption{Summary of the parameters used in \codename{} to describe the GW signal. The first column reports the symbol used to denote a parameter, the second a brief description of its physical meaning, the third its name in \codename{}, the fourth the physical units of the parameter adopted in \codename{} and the last its physical range. Parameters describing the same physical quantities, which thus have to be provided alternatively, are followed by a superscript in the first column, matching the one reported in the list in \autoref{sec:input}. $S^2$ in the $\chi_{i,c}$ stresses that the 3 components of a spin vector are not independent, but defined on a sphere, i.e. $\chi_{i,x}^2 + \chi_{i,y}^2 + \chi_{i,z}^2 \leq 1$.}
    \label{tab:input_pars}
    \vspace{-1.cm}
\end{table}

\newpage
\subsection{Waveform models}\label{sec:waveforms}
\begin{figure}[b]
\hspace{-2.cm}
    \begin{tabular}[t]{lll}
    \includegraphics[width=6.3cm]{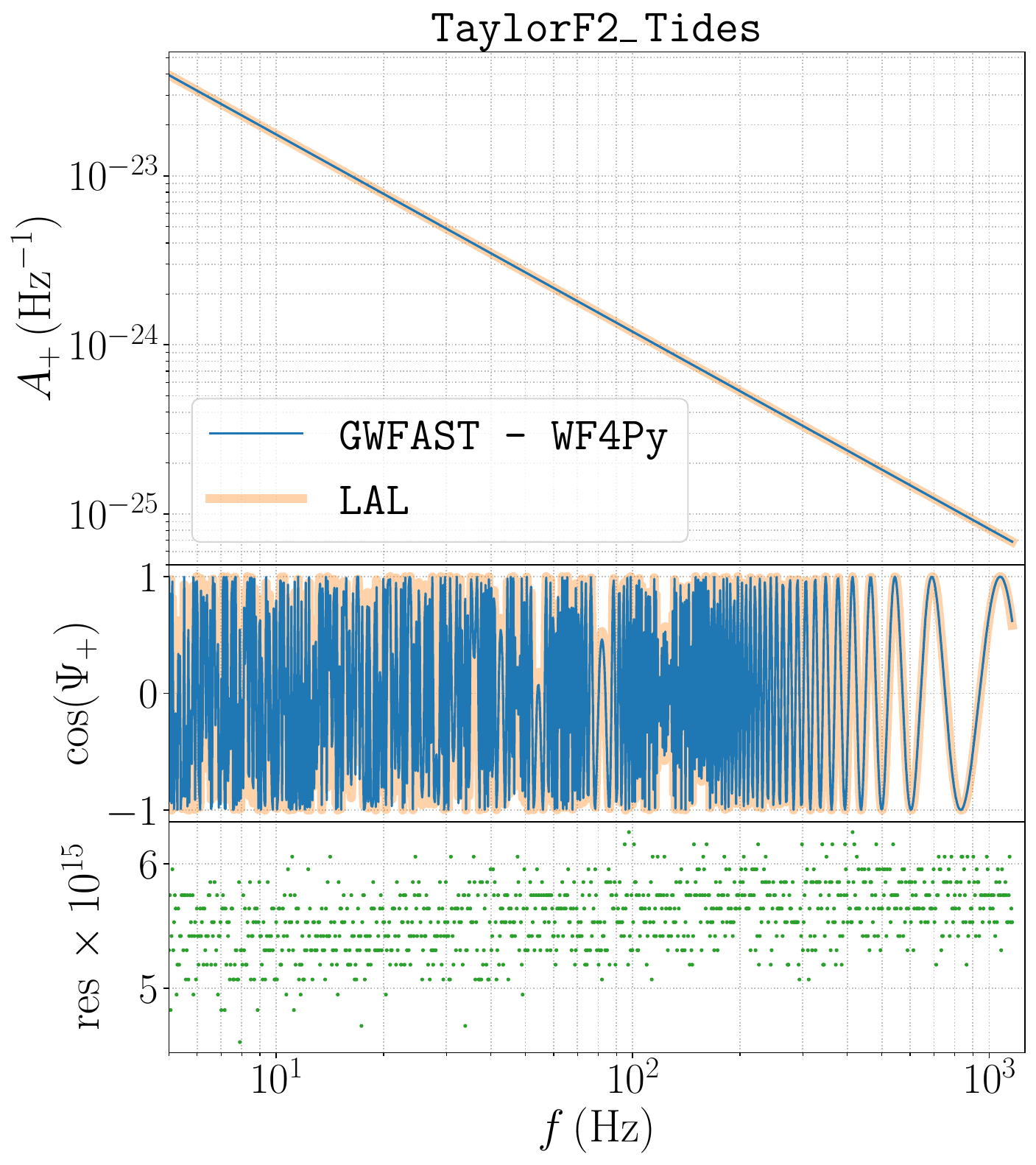} & \includegraphics[width=6.3cm]{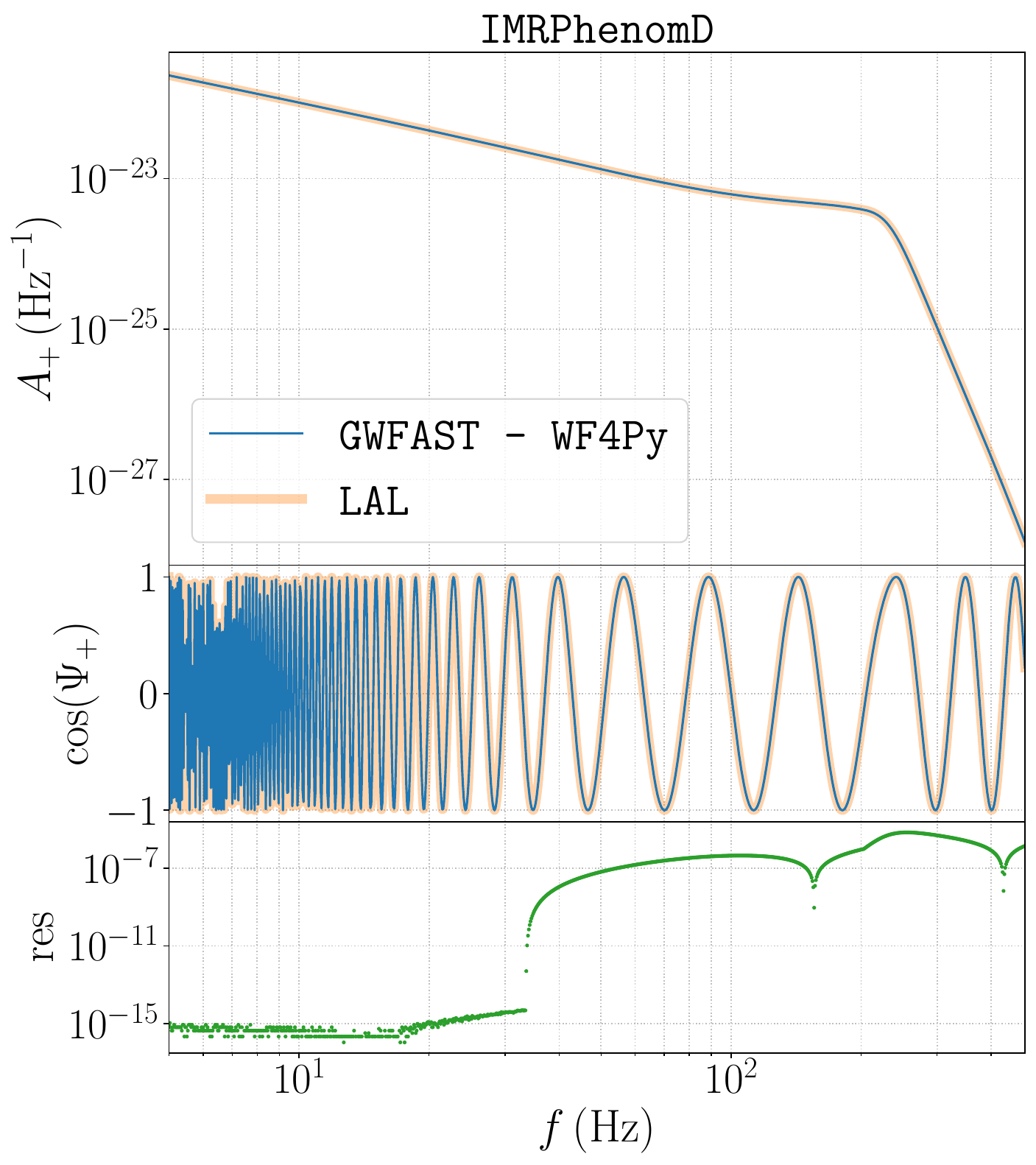} & \includegraphics[width=6.3cm]{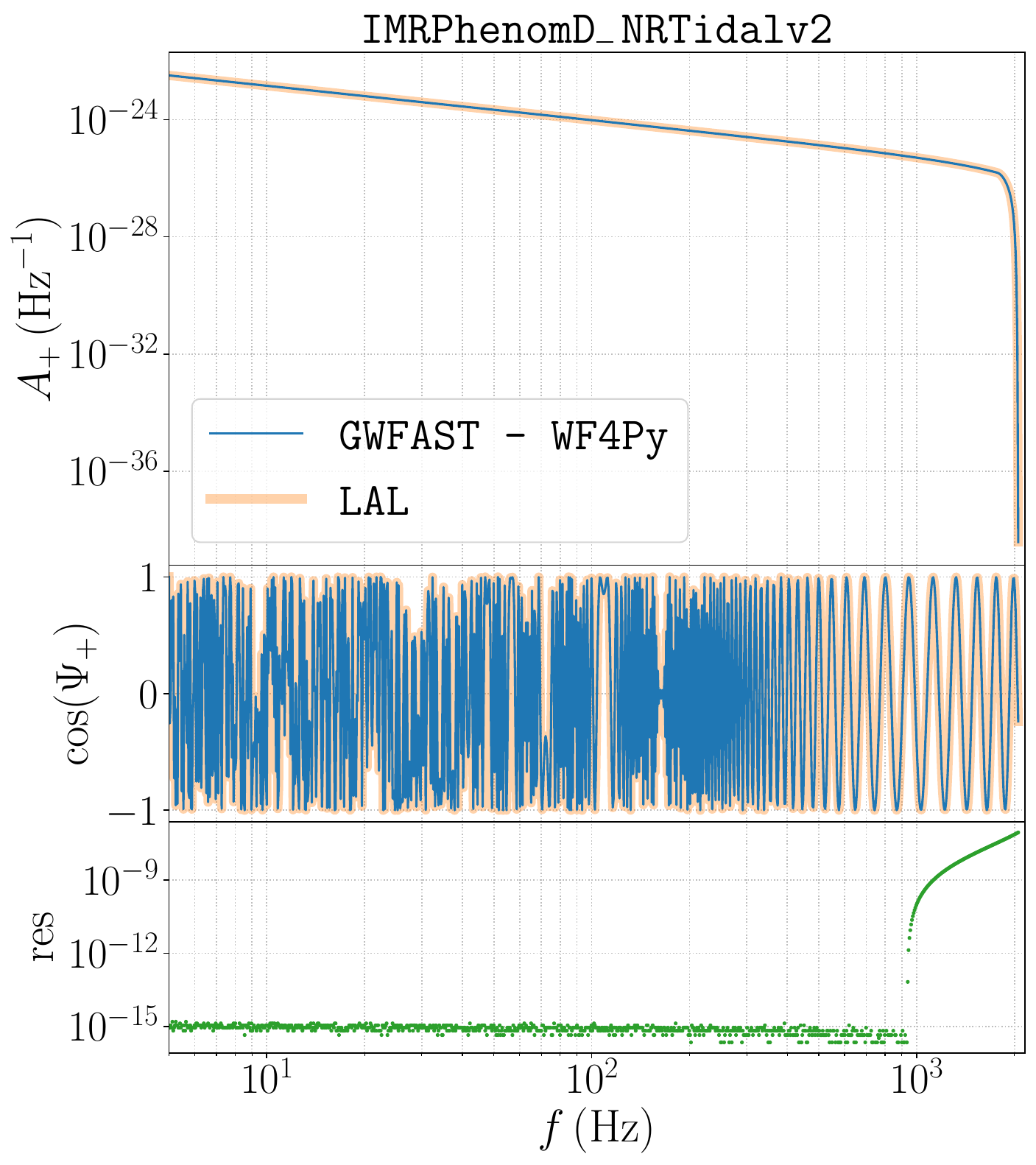} \\
    \hspace{7.5mm}{\footnotesize (a) Parameters: ${\cal M}_c = \SI{1.65}{\Msun},\ \eta=0.25,$} &  \hspace{7.5mm}{\footnotesize (b) Parameters: ${\cal M}_c = \SI{36}{\Msun},\ \eta=0.24,$} & \hspace{7.5mm}{\footnotesize (c) Parameters: ${\cal M}_c = \SI{1.32}{\Msun},\, \eta=0.25,$} \\
    \hspace{12.5mm}{\footnotesize $d_L = \SI{0.46}{\giga\parsec},\ \chi_1 = 0.05,\ \chi_2 = -0.05,$} & \hspace{12.5mm}{\footnotesize$ d_L = \SI{0.98}{\giga\parsec},\ \chi_1 = 0.8,\ \chi_2 = -0.8$.} & \hspace{12.5mm}{\footnotesize$d_L = \SI{0.46}{\giga\parsec},\ \chi_1 = 0.05,\ \chi_2 =0,$} \\
    \hspace{12.5mm}{\footnotesize $\Lambda_1 = 300,\ \Lambda_2 = 300$.} & & \hspace{12.5mm}{\footnotesize $\Lambda_1 = 500,\ \Lambda_2 = 500$.}
  \end{tabular}%
    \caption{Comparison of the waveform models \texttt{TaylorF2\_RestrictedPN} (left panel), \texttt{IMRPhenomD} (central panel) and \texttt{IMRPhenomD\_NRTidalv2} (right panel) obtained from \texttt{LAL} and \codename{} -- \wfname{} for example events. The upper and central panel of both figures show the ‘+' GW amplitude and cosine of the phase obtained with the two codes superimposed, while in the lower we report the relative difference (‘‘residual'') among the two amplitudes.}
    \label{fig:WF_compar_PhenD_normandTid}
\end{figure}
We use Fourier domain waveform models. In particular, at the time of writing the code implements some selected waveform models in \texttt{Python}, and an interface to the LIGO Algorithm Library, \texttt{LAL}, which allows to use all the waveforms available in that library.\\
In particular, the following waveform models are directly available in \codename{} in a \texttt{Python} implementation:

\begin{description}[align=left]
    \item[\texttt{TaylorF2\_RestrictedPN}] a restricted PN waveform model \citep{Buonanno:2009zt, PhysRevD.84.084037, PhysRevD.93.084054}, also with its tidal \citep{PhysRevD.89.103012} and moderate eccentric \citep{Moore:2016qxz} extensions. This is an inspiral--only waveform, but can still be used to describe signals coming from BNS mergers, whose major contribution to the SNR comes from the inspiral. There is no limitation in the parameters range, except for the eccentricity, which cannot exceed $e_0\sim0.1$ for comparable mass systems;\footnote{The reference cut frequency of the waveform is set by default to twice the binary Innermost Stable Circular Orbit frequency, $f_{\rm ISCO} = 1/(2 \pi \, 6\sqrt{6} \, G M_{\rm tot}/c^3)$. The tidal and eccentric terms can be included through the Booleans \texttt{is\_tidal} and \texttt{is\_eccentric}, respectively (see also \footref{footnote:fRef_ecc}).}
    \item[\texttt{IMRPhenomD}] a full inspiral--merger--ringdown waveform model \citep{PhysRevD.93.044006, PhysRevD.93.044007}, tuned with NR simulations, which can efficiently be used to simulate signals coming from BBH mergers, with non--precessing spins up to $|\chi_z| \sim 0.85$ and mass ratios up to $q = m_1/m_2 \sim 18$;  
    \item[\texttt{IMRPhenomD\_NRTidalv2}] tidal extension of the previous model \citep{PhysRevD.100.044003} which can be used to accurately describe signals coming from BNS mergers. It includes spin terms up to higher order and a filter to terminate the waveform after merger. The validity has been assessed for masses ranging from 1 to \SI{3}{\Msun}, spins up to $|\chi_z| \sim 0.6$ and tidal deformabilities up to $\Lambda_i\simeq5000$;
    \item[\texttt{IMRPhenomHM}] full inspiral--merger--ringdown waveform model \citep{PhysRevLett.120.161102, PhysRevD.101.103004}, which takes into account not only the quadrupole of the signal, but also the sub--dominant multipoles $(l,m) = (2,1),\ (3,2),\ (3,3),\ (4,3),\ {\rm and}\ (4,4)$, that can be particularly relevant to better describe the signal coming from BBH systems. The calibration range is the same of the \texttt{IMRPhenomD} model;
    \item[\texttt{IMRPhenomNSBH}] full inspiral--merger--ringdown waveform model \citep{PhysRevD.92.084050, PhysRevD.100.044003}, which can describe the signal coming from the merger of a NS and a BH, with mass ratios up to $q\sim 100$, also taking into account tidal effects and the impact of the possible tidal disruption of the NS.
\end{description}

\begin{figure}[t]
\hspace{-2.cm}
\begin{tabular}[t]{lll}
    \includegraphics[width=6.3cm]{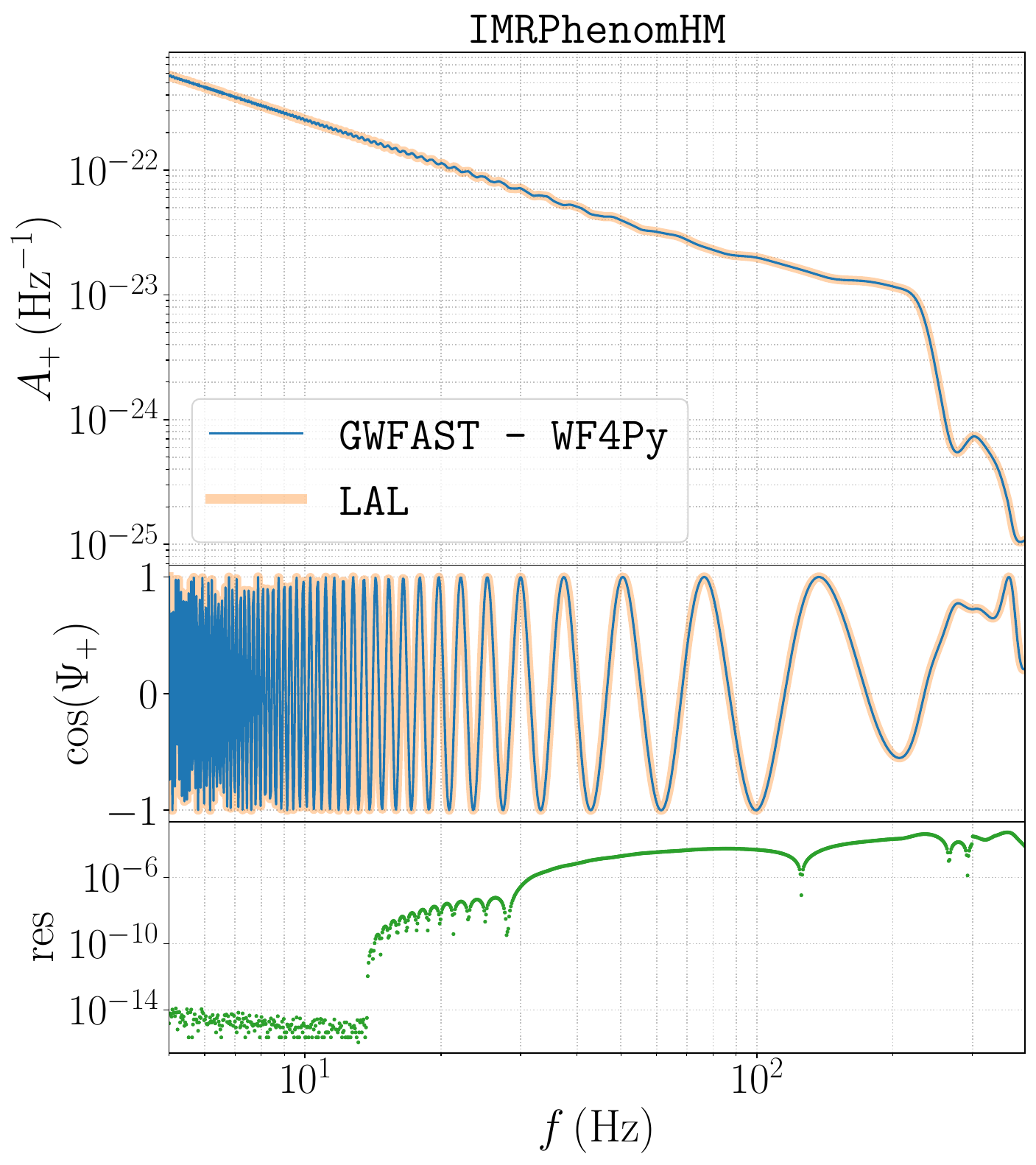} & \includegraphics[width=6.3cm]{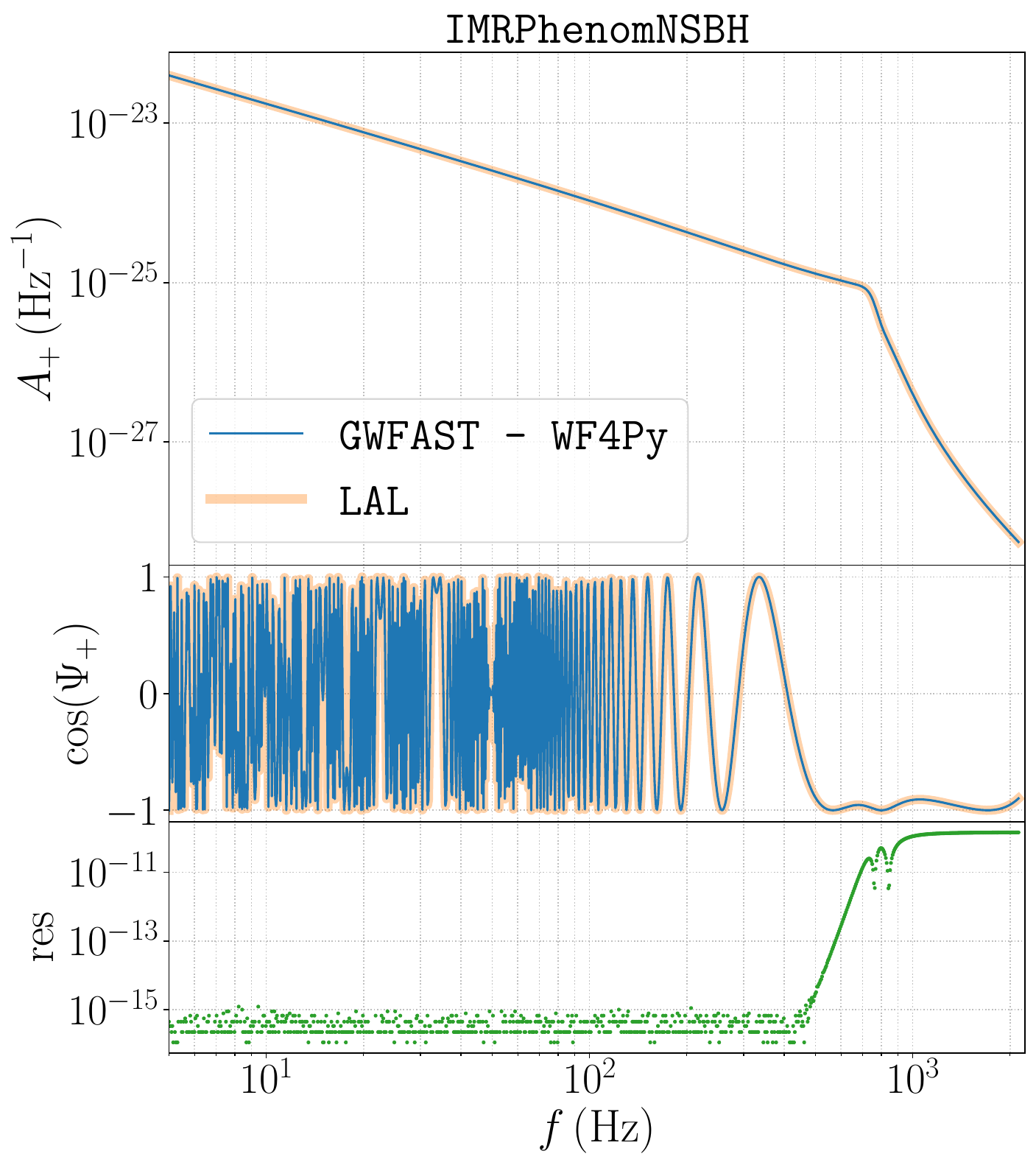} & \includegraphics[width=6.3cm]{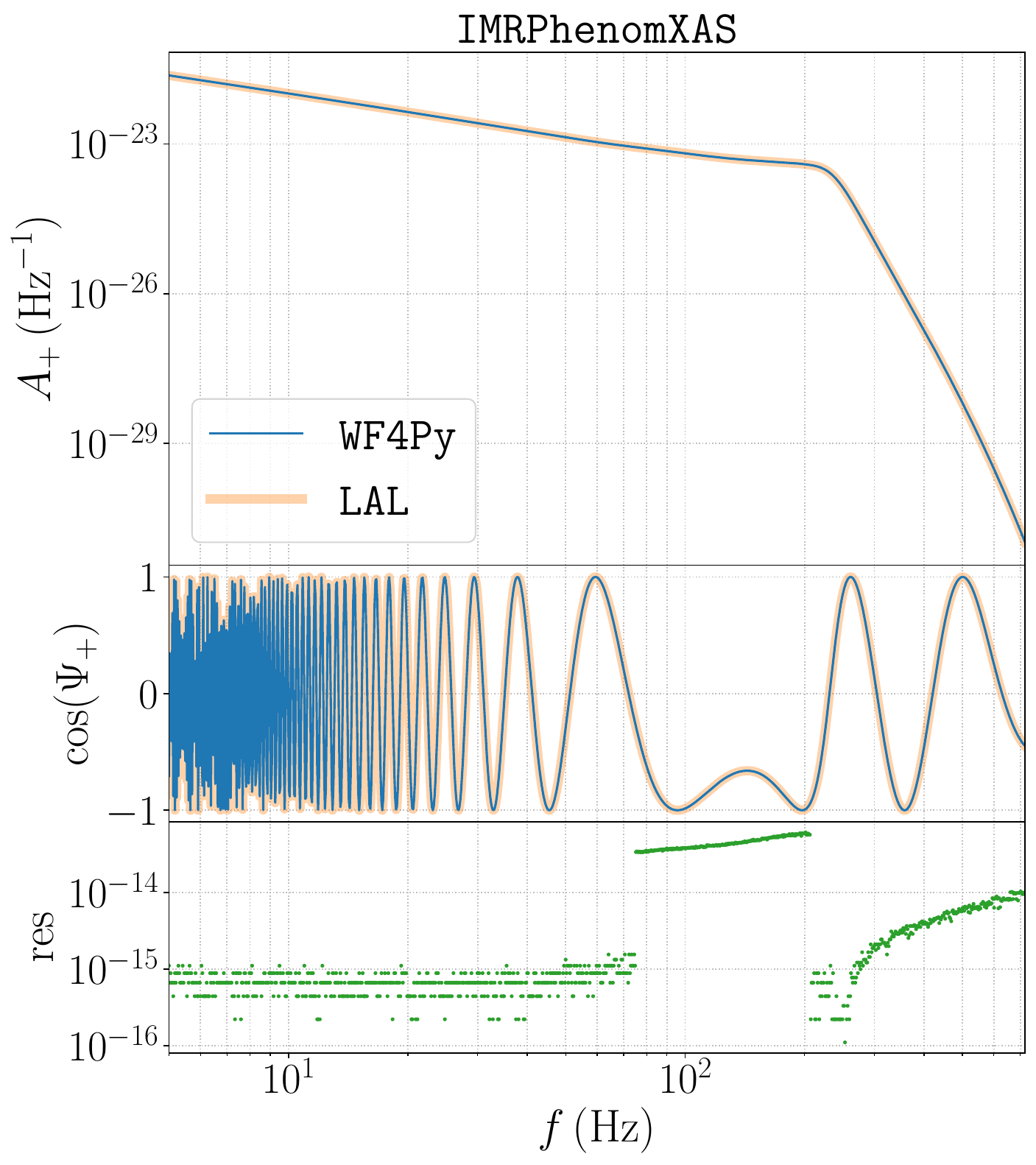} \\
    \hspace{7.5mm}{\footnotesize (a) Parameters: ${\cal M}_c = \SI{44}{\Msun},\ \eta=0.24,$} &  \hspace{7.5mm}{\footnotesize (b) Parameters: ${\cal M}_c = \SI{3.85}{\Msun},\ \eta=0.08,$} & \hspace{7.5mm}{\footnotesize (c) Parameters: ${\cal M}_c = \SI{36}{\Msun},\ \eta=0.24,$} \\
    \hspace{12.5mm}{\footnotesize $d_L = \SI{0.46}{\giga\parsec},\ \chi_1 = 0.8,\ \chi_2 = 0.8 $.} & \hspace{12.5mm}{\footnotesize$d_L =\SI{0.46}{\giga\parsec},\ \chi_1 = 0.3,\ \chi_2 =0, $} & \hspace{12.5mm}{\footnotesize$d_L = \SI{0.98}{\giga\parsec}, \ \chi_1 = 0.8,\ \chi_2 = -0.8$.} \\
    & \hspace{12.5mm}{\footnotesize $\Lambda_1 = 0,\ \Lambda_2 = 400$.} &
  \end{tabular}%
    \caption{As in \autoref{fig:WF_compar_PhenD_normandTid} for the waveform models \texttt{IMRPhenomHM} (left panel), \texttt{IMRPhenomNSBH} (central panel) and \texttt{IMRPhenomXAS} (right panel).}
    \label{fig:WF_compar_Phen_HMandNSBHandXAS}
    \vspace{-.3cm}
\end{figure}
These waveform models have been translated from their \texttt{C} implementation in the LIGO Algorithm Library, \texttt{LAL}, into a pure \texttt{Python} version.
We carefully checked that our \texttt{Python} implementation accurately reproduces the original \texttt{LAL} waveforms, as can be seen on some example events in \autoref{fig:WF_compar_PhenD_normandTid} and \ref{fig:WF_compar_Phen_HMandNSBHandXAS}.\footnote{Some bigger fluctuations arise when going closer to the merger, but this can be explained by small differences among \texttt{C} and \texttt{Python} in some interpolations needed to compute the ringdown frequency, as we prove in the following.} 

The implementation in \texttt{Python} has two advantages. First, it allows \codename{} to fully exploit the capabilities of this language to vectorize the computation on multiple events at a time, which would be impossible if we had to interact with a code written in \texttt{C} such as \texttt{LAL}. 
Second, it allows the possibility of using automatic differentiation (and in particular the library \texttt{JAX}) to compute derivatives, see \autoref{sec:signal_derivs}.

It is also possible to use the \texttt{waveform} module separately from the rest of the code. For example, in order to generate the \texttt{IMRPhenomD} waveform amplitude and phase for a given set of events it is sufficient to run the following sequence of commands
\begin{lstlisting}[caption={Calculation of waveforms in \codename{} and \wfname{}. As an illustration, we use the \texttt{IMRPhenomD} model.},label={lst:wfs_code}]
import waveform as wf

# initialise the waveform, here we choose as an example IMRPhenomD
mywf = wf.IMRPhenomD()

# compute the cut frequencies
fcut = mywf.fcut(**events)

# initialise the frequency grids from 2Hz to fcut, with 1000 points per grid
fgrids = np.geomspace(np.full(fcut.shape, 2), fcut, num=1000)

# compute the amplitude and phase, respectively
myampl = mywf.Ampl(fgrids, **events)
myphase = mywf.Phi(fgrids, **events)

\end{lstlisting}
This small piece of code shows that, being \codename{}'s waveforms written in pure \texttt{Python} and fully vectorized, our software does not have to rely on \codekeyword{for} loops over the events, as in a code interacting with \texttt{C}. Note that the order of the entries in the \texttt{events} dictionary is arbitrary.

All our waveforms also include a routine to compute the time to coalescence as a function of frequency, needed to take into account Earth's rotation in the computation of the strain, which includes terms up to 3.5\,PN order,\footnote{As in \cite{Buonanno:2009zt}, Eq. (3.8b).} called \texttt{tau\_star}, and a function to compute the cut frequency for the given waveform, so to properly build the frequency grid, called \texttt{fcut}, as seen in the above example.
Waveforms objects in \codename{} contain the attribute \texttt{ParNums} giving a dictionary of the form \texttt{\string{\codestring{name\_of\_parameter}:position\string}}, with \codestring{name\_of\_parameter} being a \codekeyword{string} with the parameter name as in \autoref{tab:input_pars} and \texttt{position} being an \codekeyword{int} corresponding to the position of the parameter in the Fisher matrix.

Apart from their implementation in \codename{}, which includes some features specific for \texttt{JAX} compatibility, we publicly release a pure \texttt{numpy} and ready--to--use version of the waveform models alone, \wfname{} \raisebox{-1pt}{\href{https://github.com/CosmoStatGW/WF4Py}{\includegraphics[width=10pt]{GitHub-Mark.pdf}}}. The syntax for using waveforms in this library is the same as in the example above. This module further implements the waveform model \texttt{IMRPhenomXAS} \citep{Pratten:2020fqn}, which is a full inspiral--merger--ringdown model tuned for the fundamental mode of BBH systems with aligned spins and mass ratios ranging from 1 to 1000, among the last to be released.\footnote{Note that the comparison of this waveform model with its LAL implementation explains the larger fluctuations in the difference between \wfname{} and LAL arising close to the merger for the other full inspiral--merger--ringdown waveforms, see \autoref{fig:WF_compar_PhenD_normandTid} and \ref{fig:WF_compar_Phen_HMandNSBHandXAS}. For \texttt{IMRPhenomXAS}, the ringdown frequency is computed from an analytical expression rather than an interpolation of a pre--computed table, and the adherence of \wfname{} to the \texttt{LAL} original is close to machine precision, as can be seen from the right panel of \autoref{fig:WF_compar_Phen_HMandNSBHandXAS}.}

Finally, all waveform models available in \texttt{LAL} can be accessed in \codename{} through the wrapper class \texttt{LAL\_WF}, which can be used as follows
\begin{lstlisting}[caption={How to use \texttt{LAL} waveforms in \codename{}. As an illustration, we use the \texttt{IMRPhenomXPHM} model.},label={lst:LAL_wfs_code}]
myLALwf = wf.LAL_WF('IMRPhenomXPHM', is_tidal=False, is_HigherModes=True, is_Precessing=True, is_eccentric=False)
\end{lstlisting}
where the first entry has to be a \codekeyword{string} containing the name of the chosen waveform as in the \texttt{LAL} library~\footnote{Note that, in case of wrong input, the code will print a list of all the available waveform models in frequency domain.} -- \codestring{IMRPhenomXPHM} in this example -- and the Booleans \texttt{is\_tidal}, \texttt{is\_HigherModes}, \texttt{is\_Precessing} and \texttt{is\_eccentric} are used to specify whether the model includes tidal effects, the contribution of higher--order harmonics, precessing spins or eccentricity (see also \footref{footnote:fRef_ecc}), respectively. 

\section{Detector response and Fisher Matrix }\label{sec:sig_net}
The core modules of \codename{} are \texttt{signal} and \texttt{network}. The former allows to model the response of a single detector to a GW signal, while the latter collects multiple detectors constituting a network. Both modules include in particular methods to compute the signal--to--noise ratio (SNR) and the FIM. \codename{} fully accounts for the motion of the Earth during the time the signal remains in the detection band, see \cite{Iacovelli:2022bbs} for a detailed discussion.\footnote{In particular, this enters in the pattern functions and through a Doppler shift in the phase, and can give a strong contribution in particular for BNS signals, that, for 3G detectors, can stay in band up to ${\cal O}(\SI{1}{\day})$.}

\subsection{Single detector}
\begin{table}[t]
    \centering\hspace{-2.5cm}
    \begin{tabular}{!{\vrule width .09em}c|c|c|c|c|c|c!{\vrule width .09em}}
    \toprule\midrule
    Detector & latitude $\lambda$ & longitude $\varphi$ & orientation $\gamma$ & arms aperture $\zeta$ & shape & name in \codename{} \\
    \midrule\midrule
    LIGO Hanford, USA &  \SI{46.5}{\degree} & \SI{-119.4}{\degree} & \SI{171}{\degree} & \SI{90}{\degree} & \codestring{L} & \codestring{H1} \\
    \midrule
    LIGO Livingston, USA &  \SI{30.6}{\degree} & \SI{-90.8}{\degree} & \SI{242.7}{\degree} & \SI{90}{\degree} & \codestring{L} & \codestring{L1} \\
    \midrule
    Virgo, Cascina, IT &  \SI{43.6}{\degree} & \SI{10.5}{\degree} & \SI{115.6}{\degree} & \SI{90}{\degree} & \codestring{L} & \codestring{Virgo}\\
    \midrule
    KAGRA, Hida, JP &  \SI{36.4}{\degree} & \SI{137.3}{\degree} & \SI{15.4}{\degree} & \SI{90}{\degree} & \codestring{L} & \codestring{KAGRA} \\
    \midrule
    LIGO India, Hingoli, IN &  \SI{19.6}{\degree} & \SI{77.0}{\degree} & \SI{287.4}{\degree} & \SI{90}{\degree} & \codestring{L} & \codestring{LIGOI} \\
    \midrule
    ET Sardinia, IT &  \SI{40.5}{\degree} & \SI{9.4}{\degree} & \SI{0}{\degree} & \SI{60}{\degree} & \codestring{T} & \codestring{ETS}\\
    \midrule
    ET Meuse--Rhine, EU &  \SI{50.7}{\degree} & \SI{5.9}{\degree} & \SI{0}{\degree} & \SI{60}{\degree} & \codestring{T} & \codestring{ETMR}\\
    \midrule
    CE1 Idaho, USA &  \SI{43.8}{\degree} & \SI{-112.8}{\degree} & \SI{-45}{\degree} & \SI{90}{\degree} & \codestring{L} & \codestring{CE1Id}\\
    \midrule
    CE2 New Mexico, USA &  \SI{33.2}{\degree} & \SI{-106.5}{\degree} & \SI{-105}{\degree} & \SI{90}{\degree} & \codestring{L} & \codestring{CE2NM}\\
    \midrule
    CE2 New South Wales, AU &  \SI{-34}{\degree} & \SI{145}{\degree} & \SI{0}{\degree} & \SI{90}{\degree} & \codestring{L} & \codestring{CE2NSW}\\
    \midrule\bottomrule
    \end{tabular}
    \caption{Summary of the positions, orientations, angle between arms, shapes and acronyms of the detectors available in \codename{}. Using user-defined configurations is straightforward (see the text). The orientation $\gamma$ denotes the angle between the bisector of the arms (the first arm in the case of a triangle) and East.}
    \label{tab:detectorsData}
\end{table}
A signal object can be initialized from the class \texttt{GWSignal} as follows:
\begin{lstlisting}[caption={Initialization of objects characterising single detectors in \codename{} for a user--specified location, orientation, shape and path to the PSD file.},label={lst:signal}]
import signal

Virgo = signal.GWSignal(mywf, psd_path= 'path/to/Virgo/psd', detector_shape = 'L', det_lat=43.6, det_long=10.5, det_xax=115.) 

LIGO_L = signal.GWSignal(mywf, psd_path= 'path/to/LIGO/L1/psd', detector_shape = 'L', det_lat=30.6, det_long=-90.8, det_xax=243.) 
\end{lstlisting}
where \texttt{det\_lat} and \texttt{det\_long} denote the latitude and longitude of the detector in degrees, \texttt{xax} the angle between the bisector of the detector arms and the east in degrees, \texttt{detector\_shape} its shape, which can be either \codestring{L} (for L--shaped detectors) or \codestring{T} (for triangular--shaped detectors). A triangular--shaped detector is defined as three co--located detectors with opening angle of \SI{60}{\degree} in a closed--loop configuration.

Other options can be passed when initialising the \texttt{GWSignal} object, in particular: the \texttt{useEarthMotion} Boolean is used to turn on and off the computation of the effect of Earth's rotation; \texttt{fmin} and \texttt{fmax} can be used to set the minimum and maximum of the frequency grid (in \si{\hertz}) and have default values \texttt{fmin\textcolor{neonpurple}{=}\textcolor{brightgreen}{2}.\textcolor{brightgreen}{0}} and \texttt{fmax\textcolor{neonpurple}{=}\codekeyword{None}} (meaning that the grid extends up to the cut frequency of the waveform); \texttt{DutyFactor} can be used to set the duty factor of the detector, i.e. the fraction of time each detector is supposed to be operational, between 0 and 1 (default is \texttt{\codekeyword{None}}, meaning no duty cycle is considered). For triangular--shaped detectors, the duty factor refers to each of the three components of the triangle separately.

The entry \texttt{psd\_path} is the path to the file containing its Amplitude Spectral Density (ASD) or Power Spectral Density (PSD) (the flag \texttt{is\_ASD} can be used to specify whether the given one is a PSD or ASD).
\codename{} contains the following publicly available ASDs in the folder \texttt{data/psds/}:
\begin{itemize}[label=--]
    \item the sensitivities from the study \cite{LIGOScientific:2016wof} (last update in January 2020), that can be found at \url{https://dcc.ligo.org/LIGO-T1500293/public}, in the folder \texttt{unofficial\_curves\_all\_dets};
    \item the representative sensitivities of the LIGO and Virgo detectors during the observing runs O1 and O2 (\url{https://dcc.ligo.org/P1800374/public/}), O3a (\url{https://dcc.ligo.org/LIGO-P2000251/public}) and O3b (estimated using \texttt{PyCBC} around the times reported in the caption of Fig. 2 of \cite{LIGOScientific:2021djp}), in the folder \texttt{LVC\_O1O2O3};
    \item the sensitivities adopted in \cite{AbbottLivingRevGWobs} for the LIGO, Virgo and KAGRA detectors during the O3, O4 and O5 runs (\url{https://dcc.ligo.org/LIGO-T2000012/public}), in the folder \texttt{observing\_scenarios\_paper};
    \item the official ET--D  sensitivity curve, from the document \texttt{ET-0000A-18.txt} (\url{https://apps.et-gw.eu/tds/?content=3&r=14065});
    \item the latest sensitivity curves for Cosmic Explorer, used in \cite{Srivastava:2022slt}, for various detector configurations (\url{https://dcc.cosmicexplorer.org/CE-T2000017/public}), in the folder \texttt{ce\_strain}.
\end{itemize} 
\codename{} also contains some pre--defined detector configurations in the module \texttt{globals}. These are listed, together with their acronyms, in \autoref{tab:detectorsData}. The locations of the current detectors are taken from \cite{Gossan:2021eqe}, while the CE sites are taken from \cite{Borhanian:2020ypi} as illustrative examples.
Pre--defined detector configurations can be easily imported from \texttt{globals}. In the following, we show how to initialise one signal corresponding to one CE at the Hanford site, with orientation $\gamma=0$:
\begin{lstlisting}[caption={Initialization of objects characterising single detectors in \codename{} using pre--defined detector configurations.},label={lst:signalPredef}]
import copy
import os
import gwfastGlobals as glob

# copy the location and orientation of Hanford
CEH_conf = copy.deepcopy(glob.detectors).pop('H1')

# set the detector PSD using the latest curve for CE1
CEH_conf['psd_path'] = os.path.join(glob.detPath, 'ce_strain', 'cosmic_explorer.txt')

# Set the orientation angle to 0
CEH_conf['xax'] = 0

# Initialise the GWSignal object
CEH = signal.GWSignal(mywf, psd_path=CEH_conf['psd_path'], detector_shape=CEH_conf['shape'], det_lat=CEH_conf['lat'], det_long=CEH_conf['long'], det_xax=CEH_conf['xax']) 

\end{lstlisting}
Any other user--defined configuration can easily be added as in \autoref{lst:signal}.
With the object of type \texttt{GWSignal} initialized, the user can easily compute all the quantities characterising signal. In particular, from the \texttt{\_PatternFunction} function it is possible to get the pattern functions of the detector, from \texttt{GWAmplitudes} it is possible to compute the ‘$+$' and ‘$\times$' amplitudes of the signal \textit{at the detector} (i.e. multiplied by the pattern functions and the spherical harmonics), while the full signal strain can be obtained through the function \texttt{GWstrain}.
\subsection{Detector networks}\label{sec:nets}
From more than one \texttt{GWSignal} objects, one can define a network, which is composed by multiple detectors. The detectors composing the network have to be inserted into a dictionary, which can then be used to initialize an object from the class \texttt{DetNet}, characterizing the network:
\begin{lstlisting}[caption={Initialisation of the network object in \codename{} from single detector objects.},label={lst:network_init}]
import network

# First collect the signal objects into a dictionary
mySignals = {'V1':Virgo, 'L1':LIGO_L, ...}

# Then initialise the network object
myNet = network.DetNet(mySignals) 
\end{lstlisting}
From both the signal and network objects it is then possible to compute the SNRs and Fisher matrices for a set of events. The \emph{matched filter} SNRs in a single detector is computed using the definition
\begin{equation}
    {\rm SNR}_i^2=4 \int_{f_{\rm min}}^{f_{\rm max}} \frac{|\tilde{h}_{(i)}(f)|^2}{S_{n,i}(f)}\dd{f}\, ,
\end{equation}
where $\tilde{h}_{(i)}(f)$ denotes the GW strain in Fourier domain at the $i$\textsuperscript{th} detector and $S_{n,i}(f)$ the noise spectral density of the $i$\textsuperscript{th} detector.
The network SNR is defined as the sum in quadrature of the single detectors' SNRs:
\begin{equation}
    {\rm SNR}^2 = \sum\nolimits_{i}  {\rm SNR}_{i}^2\, .
\end{equation}
This can be obtained as simply as
\begin{lstlisting}[caption={Computation of SNRs in \codename{} both for a single detector and for a network.},label={lst:SNR_comp}]
# If the SNRs in a single detector are needed, e.g. Virgo
SNRsOne = Virgo.SNRInteg(events)
# Instead for the network SNRs
SNRsNet = myNet.SNR(events) 
\end{lstlisting}
The output of the methods above is a \texttt{numpy} array of the same length of the number of events. 

The FIM elements for a single detector are computed from the definition
\begin{equation}
    \Gamma_{ij} = 4 \int_{0}^{\infty}{\rm d}f \frac{\partial_i\tilde{h}\partial_j\tilde{h}^*}{S_{n}(f)}\,,
\end{equation}
where $\partial_i$ denotes the derivative with respect to the parameter $i$.
The FIM for the network is obtained by summing the individual Fisher matrices (which relies on the fact that different detectors are independent\footnote{This assumption is considered valid also for the co--located components of a triangular--shaped detector. In particular, for a triangular configuration, the output for both the SNR and FIM returned by the methods of the class \texttt{GWSignal} already include the contribution of all the three independent detectors forming the triangle, with the SNR being summed in quadrature and the FIMs being added. }).
The FIM for a single detector or a network can be obtained with a single function call:
\begin{lstlisting}[caption={Computation of Fisher matrices in \codename{} both for a single detector and for a network.},label={lst:Fish_comp}]
# If the Fisher matrices for a single detector are needed, e.g. Virgo
FisherMatrsOne = Virgo.FisherMatr(events) 
# Instead to compute them for the network
FisherMatrsNet = myNet.FisherMatr(events)
\end{lstlisting}
The FIMs are returned by \codename{} in the form of \texttt{numpy} array, treated as an array of matrices in the last dimension. For example, an array of FIMs for 5 BBH events with 9 waveform parameters will have dimension $(9, 9, 5)$. In the case of a network, it might be useful to store the SNRs and Fisher matrices of the single detectors. This can be done passing to the functions \texttt{SNR} and \texttt{FisherMatr} the flag \texttt{return\_all\textcolor{neonpurple}{=}\codekeyword{True}}. In this case the output of both functions is a dictionary, with keys corresponding to the detectors (one key for each arm in the case of a triangular--shaped detector) and a key \codestring{net} for the network SNRs and FIMs.

The default parameters for which \codename{} computes the FIM, in the quasi--circular, non--precessing and non--tidal case, are, in order,\footnote{Note that the ordering of the parameters in the FIM is fixed, and does not depend on how they are sorted in the dictionary containing the events. It can be accessed through the \texttt{waveform} class used to initialise the network, through its attribute \texttt{ParNums}, as shown in \autoref{lst:get_Fish_par}.} ${\cal M}_c$ in units of \si{\Msun}, $\eta$, $d_L$ in units of \si{\giga\parsec}, $\theta, \ \phi, \ \iota, \ \psi, \ t_c$ in units of seconds, $\Phi_c, \ {\rm and} \ \chi_s, \ \chi_a$.
In the case of precessing spins, the FIM is computed for the full set $\chi_{1,z},\ \chi_{2,z},\ \chi_{1,x},\ \chi_{2,x},\ \chi_{1,y},\ \chi_{2,y}$ in this order, and, in the BNS and NSBH case, the tidal parameters $\tilde{\Lambda}$ and $\delta\tilde{\Lambda}$ are also included.
In the eccentric case, also the parameter $e_0$ is included, and appears in the Fisher after both spins and tidal parameters. We chose to use the combinations $(\chi_s,\, \chi_a)$ instead of $(\chi_{1,z},\, \chi_{2,z})$ in the non--precessing case so to have two orthogonal parameters, but the FIM can be as well computed in terms of the latter quantities passing the flag \texttt{use\_chi1chi2\textcolor{neonpurple}{=}\codekeyword{True}} to the \texttt{FisherMatr} function. The choice of the combination $(\tilde{\Lambda},\, \delta\tilde{\Lambda})$ in place of $(\Lambda_1,\, \Lambda_2)$ is due to the fact that the parameter $\tilde{\Lambda}$ is much better constrained than the two dimensionless tidal deformabilities separately, being the combination entering at 5\,PN order in the inspiral signal. It is also possible to compute the FIM in terms of the combination $(m_1,\, m_2)$, i.e. the two component redshifted masses, in units of \si{\Msun}, instead of $(M_c,\,\eta)$, by passing to the \texttt{FisherMatr} function the flag \texttt{use\_m1m2\textcolor{neonpurple}{=}\codekeyword{True}}. Finally, if the contribution of precessing spins is included, setting the flag \texttt{use\_prec\_ang\textcolor{neonpurple}{=}\codekeyword{True}}, instead of $\iota$ and $\chi_{i,c}$, the FIM will be computed in terms of the parameters $\theta_{JN},\ \chi_1,\ \chi_2,\ \theta_{s,1},\ \theta_{s,2},\ \phi_{JL},\ \phi_{1,2}$,  which are more commonly used in the context of parameter estimation of GW events.

As an example, to access the values of the $(d_L, \, d_L)$ elements of the FIM for all the events in the dictionary, the user just has to run
\begin{lstlisting}[caption={How to access specific FIM elements in \codename{}.}, label={lst:get_Fish_par}]
# The parameters are contained in a dictionary in the waveform class
pars = mywf.ParNums
print(FisherMatrsNet[pars['dL'], pars['dL'], : ])
\end{lstlisting}
Both the classes \texttt{GWSignal} and the \texttt{DetNet} also include a function to compute the optimal coordinates for a signal to be seen by the considered detectors (i.e. the location corresponding to the maximum SNR), as a function of the time of coalescence. This is obtained by maximizing the pattern functions, and can be accessed as

\begin{lstlisting}[caption={How to compute the optimal location of a binary for a network of detectors in \codename{} at ${\rm GMST} = \SI{0}{\day}$.}, label={lst:optimal_loc}]
best_theta, best_phi = myNet.optimal_location(0)
\end{lstlisting} 
where the time can be provided both as a GMST or as a GPS time, setting the Boolean \texttt{is\_tGPS\textcolor{neonpurple}{=}\codekeyword{True}}. The syntax to compute the optimal location is equivalent for an object of type \texttt{GWSignal}.\footnote{Notice that, however, the estimation provided by this function in the case of a network is appropriate only if the detectors have similar characteristics (i.e. PSDs and shape). It is in fact obtained by maximizing the sum in quadrature of the pattern functions, rather than of the full SNRs, which depends not only on the location of the system, but also on its parameters (which determine the merger frequency), and the detectors' sensitivity curve. Consider e.g. a network consisting of two detectors: if one of them has better capabilities for observing low mass systems (i.e. a lower PSD with respect to the other at high frequencies) and the other for high mass systems (i.e. a lower PSD at low frequencies), a higher SNR will be obtained closer to the optimal location of the former for lighter binaries, and closer to the best location of the latter for heavier ones. Thus, in this case, to estimate the location corresponding to the highest SNR, as a function not only of time, but also of the other parameters of the binary, one has either to perform sampling or maximize the full network SNR for each choice of the binary intrinsic parameters (see e.g. Sect. 2.2 of \cite{Schutz:2011tw}).}
\subsection{Signal derivatives}\label{sec:signal_derivs}
The computation of the derivatives of the signal with respect to its parameters is the key ingredient of a Fisher code. 
In its pure \texttt{Python} version,\codename{} is based on \textit{automatic differentiation} (AD) \citep{DBLP:journals/corr/abs-1811-05031} as implemented in the library \texttt{JAX} \citep{jax2018github}. Differently from finite differences techniques, automatic differentiation exploits the fact that each code function, however complex, is built up from elementary arithmetical operations and simple functions, whose derivatives are well known, thus, by applying the chain--rule repeatedly, it is possible to automatically compute derivatives of arbitrary order near machine precision, with a number of operations comparable to the original function's one. Having a properly written pure \texttt{Python} code, which is a fundamental requirement for this technique to work, it is possible to apply AD to get a fast and accurate evaluation of the GW strain derivatives, in a semi--analytic way, despite the complexity of the function, and for multiple events at a time.

\begin{figure}[t]
    \centering
    \includegraphics[width=.9\textwidth]{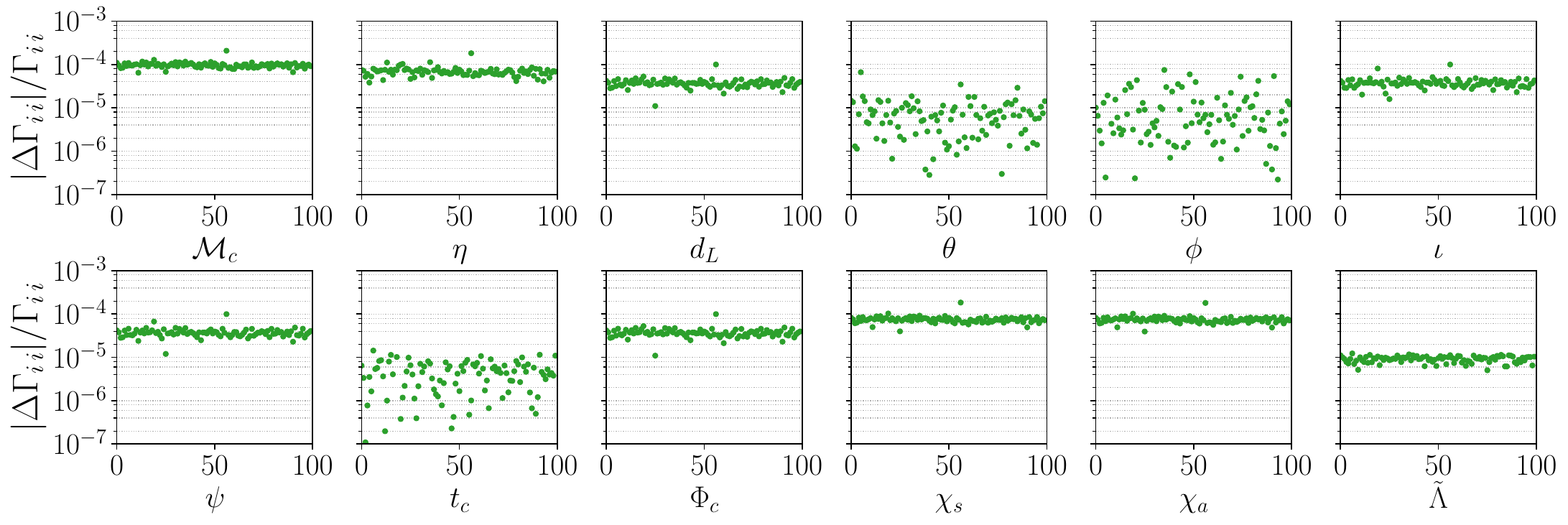}
    \caption{Relative difference of the Fisher matrices diagonal elements computed on a sample of 100 events with \codename{} and an independent \texttt{Wolfram Mathematica} code. The labels of the $x$--axes refer to the parameters whose diagonal elements are plotted.}
    \label{fig:Comparison_Fish_Math}
\end{figure}

We tested the reliability of \texttt{JAX} by comparing the Fisher matrices obtained using the \texttt{TaylorF2\_RestrictedPN} waveform model with \codename{} and an independent code written in \texttt{Wolfram Mathematica}, capable of computing derivatives analytically with respect to all parameters. The results for the relative differences of the diagonal elements of the Fisher matrices computed on a sample of 100 events are shown in \autoref{fig:Comparison_Fish_Math}, from which it is possible to see the excellent agreement among the two codes. This test acts on three levels: it proves the absence of bugs, given that the two codes were developed independently, it shows the good behaviour of \texttt{JAX} AD against an actual analytical computation, it verifies that integration is correctly performed, given the consistency of the results obtained with two different programming languages, having completely different integration routines. 

\codename{} also allows the user to compute analytically the derivatives of the signal with respect to many of the parameters, namely $d_L,\ \theta,\ \phi,\ \iota,\ \psi,\ t_c,\ {\rm and}\ \Phi_c$ to further speed up the calculation.\footnote{If the waveform model contains the contribution of sub--dominant modes or precessing spins, the dependence on the parameter $\iota$ is non trivial and linked to the modes considered, we thus do not compute the corresponding derivative analytically in this case.} This can be done passing the flag \texttt{computeAnalyticalDeriv\textcolor{neonpurple}{=}\codekeyword{True}} to the function \texttt{FisherMatr}.
We checked that, for these parameters, the analytical results and the result obtained by \texttt{JAX} agree at machine precision, i.e. $10^{-15}$.
Finally, the FIM for a triangular detector is computed by using the fact that, for a closed configuration, the sum of the signals is identically zero for geometrical reasons \citep{Freise_2009}, which further reduces the computational time by 1/3.\footnote{We also checked explicitly that the derivatives of the signal in the three arms of a triangle configuration is vanishing (up to machine precision) when computed with \codename{}.}
When using instead waveforms coming from \texttt{LAL}, which are written in \texttt{C},  \codename{} will compute derivatives using the library \texttt{numdifftools},\footnote{\url{https://pypi.org/project/numdifftools/}.} which relies on finite differences techniques. In this case, the computation is performed using the \emph{central differencing scheme} with an adaptive computation of the step size. Both these choices can be controlled though the arguments \texttt{methodNDT} and \texttt{stepNDT}, respectively.
The finite difference computation can be used alternatively to automatic differentiation also when exploiting \texttt{Python} waveforms, passing the flag \texttt{computeDerivFinDiff\textcolor{neonpurple}{=}\codekeyword{True}} to the function \texttt{FisherMatr}. Note that, also when using finite differences techniques, derivatives with respect to the parameters $d_L,\ \theta,\ \phi,\ \iota,\ \psi,\ t_c,\ {\rm and}\ \Phi_c$ can be performed analytically.

\section{Covariance matrix}\label{sec:CovMatr}
In the limit of linear signal (or large SNR) and in presence of Gaussian noise the inverse of the FIM gives the covariance of the Bayesian posterior probability distribution of the true waveform parameters for a given experiment, assuming a flat prior. The inversion of the matrix can become problematic if the condition number (i.e. the ratio of absolute values of the largest to smallest eigenvalues) is larger than the inverse machine precision. This is the case for highly correlated or nearly degenerate combinations of parameters. In this case, the linear signal approximation might break down in some regions of the likelihood surface \citep{Vallisneri:2007ev}. 
\subsection{Fisher matrix inversion}
Tools for obtaining the covariance and analyzing the reliability of the inversion are contained in the module \texttt{fisherTools}.
Note that all the functions described here assume that the input FIM is an array of matrices in the last dimension, as described in \autoref{sec:nets}
The conditioning of the FIM can be checked in \codename{} via the function \texttt{CheckFisher}, which returns the eigenvalues, eigenvectors, and condition number of the matrix.
The inversion of the FIM to yield the covariance is done with the function \texttt{CovMatr}, as 
\begin{lstlisting}[caption={Computation of covariance matrices from Fisher matrices in \codename{}.},label={lst:Cov_comp}]
import fisherTools as fTools

CovMatrsNet, inversion_errors = fTools.CovMatr(FisherMatrsNet)
\end{lstlisting}
By default, each row and column is normalized to the square root of the diagonal of the FIM before inversion, so that the resulting matrix has adimensional entries with ones on the diagonal and the remaining elements in the interval $[-1,\, 1]$~\citep{Harms:2022ymm}.\footnote{This transformation is not applied in case the matrix has a zero element on the diagonal.}
The inverse transformation is applied after inversion to yield the inverse of the original matrix.
\codename{} also implements a variety of possibilities for the technique used to find the inverse. The \texttt{Python} library for precision arithmetic \texttt{mpmath} is used for the inversion. The inversion is not performed if the condition number is larger than a threshold that can be specified by the user via the argument \texttt{condNumbMax}. Its default value is $10^{50}$ (so the code will try to invert every matrix irrespective of the conditioning). 
The available possibilities are listed below, and can be specified by the argument \texttt{invMethodIn}:
\begin{description}
    \item[\codestring{inv}] inverse computed by \texttt{mpmath};
    \item[\codestring{cho}] inverse computed by means of the Cholesky decomposition, i.e. the (Hermitian, positive--definite) FIM is expressed as a product of a lower triangular matrix and its conjugate transpose, and the latter is inverted. This is the default option in \codename{};\footnote{Note that in some case, the FIM may be not positive--definite due to the presence of very small eigenvalues that can assume negative values due to numerical fluctuations, in which case the Cholesky decomposition cannot be found. For those matrices, \codename{} resorts by default to a SVD for the inversion, but the error on the inversion in those cases should be carefully checked and the reliability of the validity of the Fisher approximation may be poor.}
    \item[\codestring{svd}] the singular--value decomposition (SVD) of the FIM is used to invert the matrix. In this case, there is the additional option of truncating the smallest singular values to the minimum allowed numerical precision, that can help regularizing badly conditioned matrices. This can be required by setting the Boolean \texttt{truncate\textcolor{neonpurple}{=}\codekeyword{True}}. In this case, for each singular value $s$, if the ratio of its absolute value to the absolute value of the largest singular value, $\max{s_i}$, is smaller than a threshold $\lambda$, the singular value $s$ is replaced with $\lambda \times \max{(s_i)}$. The value of the threshold $\lambda$ can be specified with the argument \texttt{svals\_thresh} which is set by default to $10^{15}$;
    \item[\codestring{svd\_reg}] the singular--value decomposition (SVD) of the FIM is used to invert the matrix, and eigenvalues smaller than the threshold specified by the argument \texttt{svals\_thresh} are not included in the inversion. This ensures that the error on badly constrained parameters is not propagated to the other ones~\citep{Harms:2022ymm}. However, it might result in underestimating the uncertainty for parameters whose eigenvalues are excluded, and the effect should be carefully checked.
    \item[\codestring{lu}] inversion is done by means of the Lower--Upper (LU) decomposition, i.e. the factorization of the FIM into the product of one lower triangular matrix and one upper triangular matrix. This can be a useful option since, as for the Cholesky decomposition, the inversion of a triangular matrix is easier than the one of a full matrix. Differently from the Cholesky decomposition, however, the original matrix does not have to be hermitian and positive--definite, which can make this method more stable against numerical noise for badly--conditioned matrices.
\end{description}

The error on the inversion is computed in \codename{} by the function \texttt{compute\_inversion\_error} with the definition $\epsilon = {|| \Gamma \cdot \Gamma^{-1} - \mathbb{1} ||}_{\rm max} = {\rm max}_{ij} |(\Gamma \cdot \Gamma^{-1} - \mathbb{1} )_{ij}| $, where $\mathbb{1}$ denotes the identity matrix, $\Gamma$ the FIM and $\Gamma^{-1}$ its inverse as computed by the code 
\begin{lstlisting}[caption={Computation of the inversion errors in \codename{}.},  label={lst:invErr_comp}]
invErrs = fTools.compute_inversion_error(FisherMatrsNet, CovMatrsNet)
\end{lstlisting}

Two other utilities to check the quality of the inversion are available in \codename{} in the module \texttt{fisherTools}. The function \texttt{check\_covariance} computes the inversion error, and prints the difference between $\Gamma \cdot \Gamma^{-1}$ and the identity on the diagonal, and the off--diagonal elements of $\Gamma \cdot \Gamma^{-1}$ exceeding a given threshold specified with the argument \texttt{tol}. Secondly, the function \texttt{perturb\_Fisher} adds random perturbations to the FIM to a specified decimal (given by the argument \texttt{eps}, whose default is \num{e-10}), and checks if the inversion remains stable.

While the squared root of the diagonal elements of the covariance matrix give the expected marginalized $1\sigma$ errors on the parameters, a useful metric for GW parameter estimation is the sky localization region at some given confidence level. This is computed by \citep{Barack:2003fp, Wen:2010cr}
\begin{equation}
    \Delta\Omega_{{\rm X}\%} = -2\pi |\sin\theta|\stopsqrt{\left(\Gamma^{-1}\right)_{\theta\theta}\, \left(\Gamma^{-1}\right)_{\phi\phi} - \left(\Gamma^{-1}\right)_{\theta\phi}^2}\ \ln{\left(1 - {{\rm X}}/{100}\right)}\,.
\end{equation}
The function \texttt{compute\_localization\_region} computes the sky localization region, in square degrees or steradian, according to the previous definition. The desired units can be specified through the \texttt{units} key, which can have values \codestring{SqDeg} and \codestring{Sterad}, and the confidence level is specified by the optional argument \texttt{perc\_level} (the default $90\%$). An example of usage is presented in \autoref{lst:Cov_comp_marg}.

\subsection{Manipulating the Fisher and covariance matrices}

The Fisher approach allows to treat straightforwardly some common situations encountered in parameter estimation, which we summarize here together with a description of their implementation in \codename{}. All functions described in the following belong to the module \texttt{fisherTools}:

\begin{itemize}[label=--]
    \item In order to \emph{fix some parameters to their fiducial values}, one has to remove from the FIM (before inverting it) the corresponding rows and columns. This is done with the function \texttt{fixParams}, which takes the following arguments (in the order they are listed here): the original matrix, the dictionary specifying the position of each parameter in the FIM (accessible from the waveform object, see the end of \autoref{sec:waveforms} for an explanation), and a \textcolor{brightgreen}{\texttt{list}} of \codekeyword{string} with names of the parameters to be fixed, with the same names as in \autoref{tab:input_pars}. The function returns the new matrix, and a dictionary of the same form of the input dictionary, with the keys corresponding to the fixed parameters removed, and the remaining rescaled;
    \item In order to \emph{add a Gaussian prior on some parameters}, one has to add to the FIM a prior matrix $P_{ij}$ corresponding to the inverse covariance of the prior. For the moment, \codename{} supports the addition of a diagonal prior matrix.
    This can be done with the function \texttt{addPrior}, which takes as input the original matrix, a list of values to be added on the diagonal of the Fisher (representing thus the inverse covariance of the prior on the corresponding parameter), the dictionary specifying the position of each parameter in the FIM, and the list of names of parameters on which the prior should be added;
    \item In order to \emph{marginalize over some parameters}, one has to remove from the covariance matrix (after the inversion of the FIM) the corresponding rows and columns. This can be done again with the function \texttt{fixParams} described in the first point.
\end{itemize}
\begin{lstlisting}[caption={Example of manipulations of the Fisher matrix: fix the spins to their fiducial values, add a Gaussian prior on the angles with standard deviation $2\pi$, compute the corresponding covariance, compute the forecasted $90\%$ localization region in square degrees.},label={lst:Cov_comp_marg}]
# Fix spins to their fiducial values
FisherMatrsNet_fix_spins, pars_nospin = fTools.fixParams(FisherMatrsNet, pars, ['chi1z', 'chi2z'])

# Add Gaussian prior on theta, phi, iota, psi, phicoal
angles = ['theta', 'phi', 'iota', 'psi', 'Phicoal']
priors_vals = np.repeat(1/(2*np.pi**2), len(angles))
FisherMatrsNet_fix_spins_prior = fTools.addPrior(FisherMatrsNet_fix_spins, priors_vals, pars_nospin, angles)

# Invert the new FIM
CovMatrsNet_fix_spins_prior, inversion_errors_fix_spins_prior = fTools.CovMatr( FisherMatrsNet_fix_spins_prior)

# Compute 90% localization area in square degrees
sky_loc = fTools.compute_localization_region(CovMatrsNet_fix_spins_prior, pars_nospin, events['theta'], perc_level=90, units='SqDeg')
\end{lstlisting}

\section{Running in parallel}\label{sec:parallel_run}

Besides the accuracy in the computation of the derivatives, the main advantage of the use of \texttt{JAX} \citep{jax2018github} is that it allows us to vectorize the calculation of the FIM. The typical usage of a code as \codename{} consists in forecasting parameter estimation capabilities for large catalogs of sources, which is  clearly a parallel problem. \texttt{JAX} and \codename{} allow us to vectorize the calculation \emph{even on a single CPU}, which can be used in combination with parallelization routines. \codename{} includes the executable \texttt{calculate\_forecasts\_from\_catalog.py} that implements such parallelization and is ready to use both on single machines and on clusters. 

\begin{figure}[t]
    \centering
     \subfloat{
    \includegraphics[width=8cm]{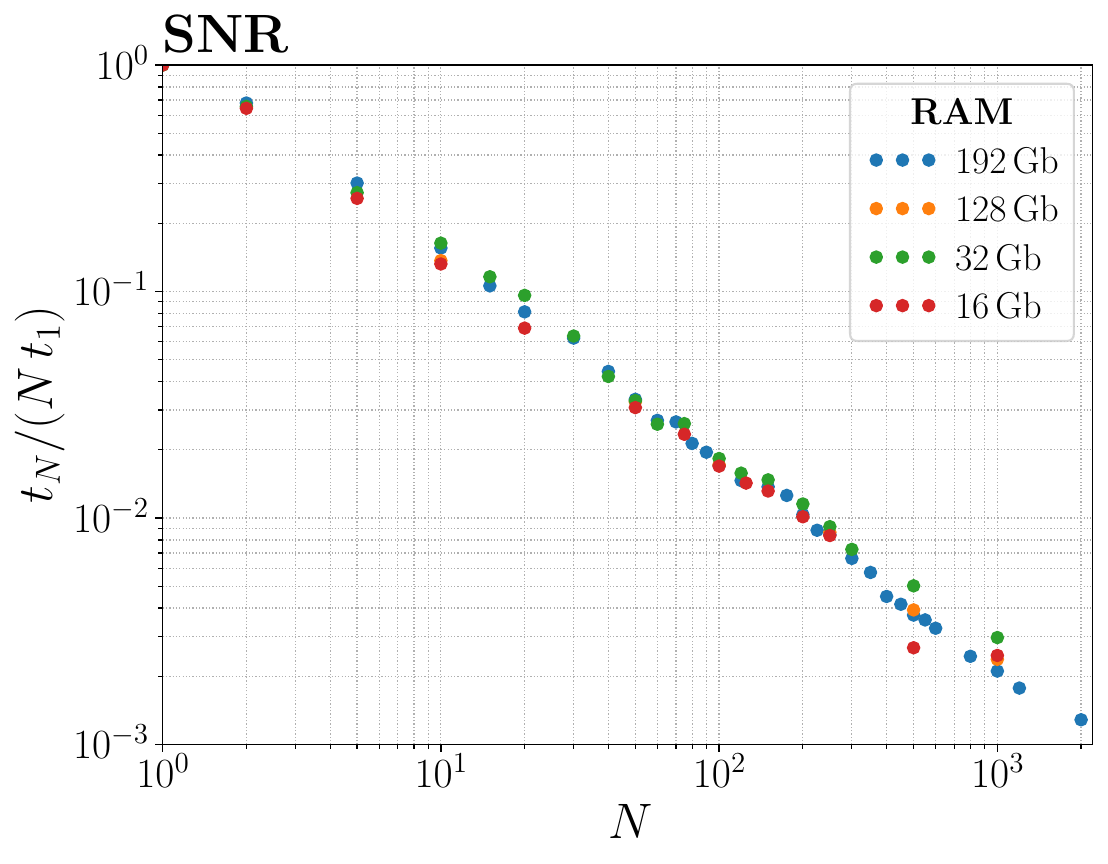}
    \label{fig:Speed_test_SNR}
    }
    \qquad
    \subfloat{
    \includegraphics[width=8cm]{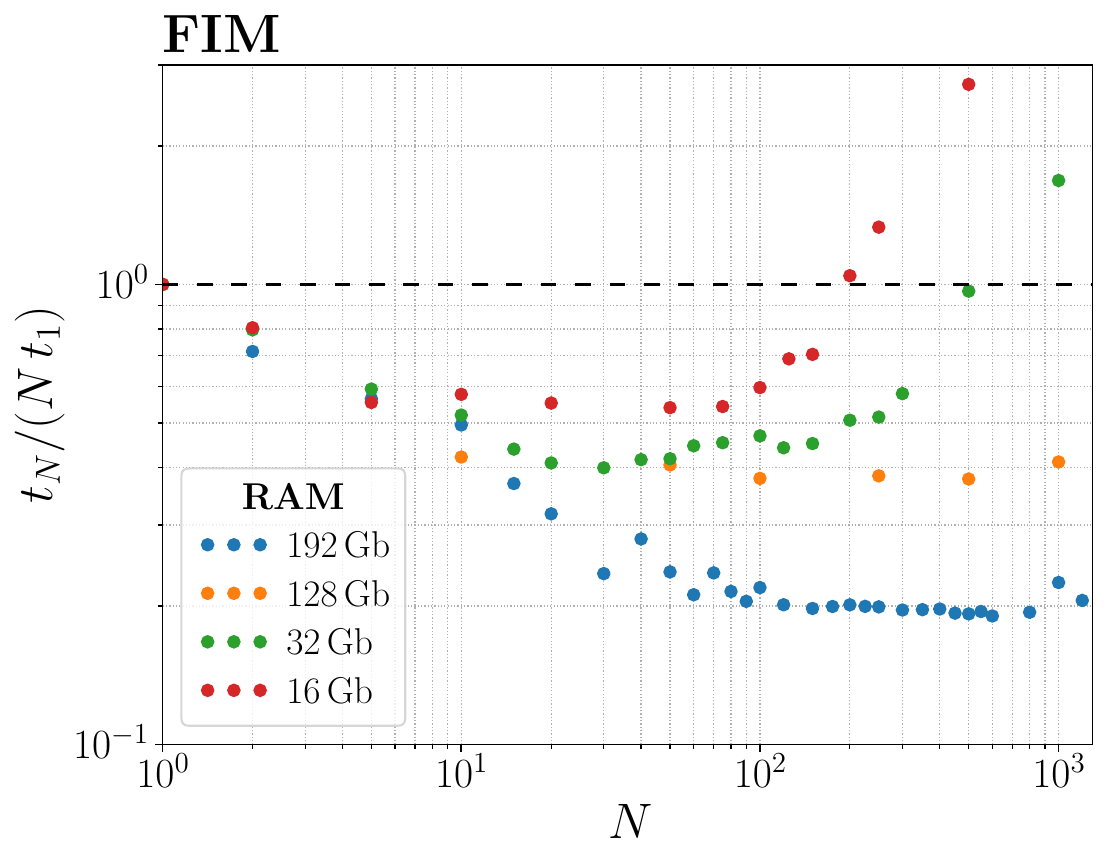}
    \label{fig:Speed_test_Fish}
    }
    \caption{Ratio of the time needed to compute SNRs (left) and Fisher matrices (right) vectorizing on $N$ events, and the time needed to perform the computation serially, with a \codekeyword{for} loop (equivalent to $N$ times the time needed for a single evaluation). The different colors refer to the results obtained using machines with different characteristics, as reported in the legend.}
    \label{fig:Speed_test}
\end{figure}

A catalog has to be stored in \texttt{.h5} format in the folder \texttt{data/}. A function to save a catalog of the form given in \autoref{lst:events_dict_init} is included in the module \texttt{gwfastUtils}:
\begin{lstlisting}[caption={How to save an event catalog in \codename{}.},label={lst:save_cat}, literate={5}{{{5}}}1]
from gwfastUtils import save_data

# save events in .h5 format
save_data('file_name.h5', events)
\end{lstlisting}

\texttt{calculate\_forecasts\_from\_catalog.py} divides the events in the catalog into batches of size specified by the user with the option \texttt{\textcolor{brightgreen}{--}batch\_size}, and splits the calculations assigning a given number of batches to each parallel process. The number of processes is controlled by the option \texttt{\textcolor{brightgreen}{--}npools}. Events in a single batch are computed in vectorized form for each process, which results effectively in a gain of speed of a factor that can be at most equal to the batch size with respect to a non--vectorized implementation. \texttt{calculate\_forecasts\_from\_catalog.py} allows both the use of \texttt{multiprocessing} \citep{2012arXiv1202.1056M} on a single machine and the use of \texttt{MPI} \citep{9439927} on clusters. The usage is as follows:
\begin{lstlisting}[caption={How to run \codename{} on a catalog of events through the script \texttt{calculate\_forecasts\_from\_catalog.py}.},label={lst:run_parallel}, language=bash, literate={-}{{{\color{brightgreen}-}}}1
{4}{{{{4}}}}1
{>}{{{\color{brightgreen}>}}}1]
> mkdir my_results
> python calculate_forecasts_from_catalog.py --fout=my_results --fname_obs FNAME_OBS      [--wf_model WF_MODEL] [--batch_size BATCH_SIZE]
    [--npools NPOOLS] [--snr_th SNR_TH] [--idx_in IDX_IN] [--idx_f IDX_F] [--fmin FMIN] 
    [--fmax FMAX] [--compute_fisher COMPUTE_FISHER] [--net NET [NET ...]] [--rot ROT] 
    [--netfile NETFILE] [--psds PSDS [PSDS ...]] [--mpi MPI] [--duty_factor DUTY_FACTOR]
    [--params_fix PARAMS_FIX [PARAMS_FIX ...]] [--lalargs LALARGS [LALARGS ...]]
    [--return_all RETURN_ALL] [--seeds SEEDS [SEEDS ...]]
\end{lstlisting}
The options are as follows:
\begin{description}
    \item[\texttt{fout}] \codekeyword{string}; path to output folder, which has to exist before the script is launched;
    \item[\texttt{fname\_obs}] \codekeyword{string}; name of the file containing the catalog \emph{without} the extension \texttt{.h5};
    \item[\texttt{wf\_model}] \codekeyword{string}; name of the waveform model, default is \codestring{tf2}. Options are: \codestring{tf2}, \codestring{tf2\_tidal}, \codestring{tf2\_ecc}, \codestring{IMRPhenomD}, \codestring{IMRPhenomD\_NRTidalv2}, \codestring{IMRPhenomHM}, \codestring{IMRPhenomNSBH}. It is also possible to choose all the other waveform models available in \texttt{LAL}, by passing \codestring{LAL-wfname}, where \texttt{wfname} is the name of the chosen waveform in \texttt{LAL}, e.g. \codestring{LAL-IMRPhenomXPHM};
    \item[\texttt{batch\_size}] \codekeyword{int}, default is \textcolor{brightgreen}{\texttt{1}}; size of the batch to be computed in vectorized form on each process;
    \item[\texttt{npools}] \codekeyword{int}, default is \textcolor{brightgreen}{\texttt{1}}; number of parallel processes;
    \item[\texttt{snr\_th}] \codekeyword{float}, default is \textcolor{brightgreen}{\texttt{12.}}; threshold value for the SNR to consider the event detectable. FIMs are computed only for events with SNR exceeding this value;
    \item[\texttt{idx\_in}] \codekeyword{int}, default is \textcolor{brightgreen}{\texttt{0}}; index of the event in the catalog from which to start the calculation;
    \item[\texttt{idx\_f}] \codekeyword{int}, default is \textcolor{brightgreen}{\texttt{-1}} (meaning all events); index of the event in the catalog from which to end the calculation;
    \item[\texttt{fmin}] \codekeyword{float}, default is \textcolor{brightgreen}{\texttt{2.}}; minimum frequency of the grid in \SI{}{Hz};
    \item[\texttt{fmax}] \codekeyword{float}, default is \codekeyword{None}; maximum frequency in \SI{}{Hz}. If not specified, coincides with the cut frequency of the waveform;
    \item[\texttt{compute\_fisher}] \codekeyword{int}, default is \textcolor{brightgreen}{\texttt{1}}; if \textcolor{brightgreen}{\texttt{0}}, only SNRs are computed, if \textcolor{brightgreen}{\texttt{1}} the code also computes FIMs; 
    \item[\texttt{net}] \textcolor{brightgreen}{\texttt{list}} of \codekeyword{string}, default is \texttt{[}\codestring{ETS}\texttt{]}; the network of detectors chosen. Pre--defined configurations are passed using the names in \autoref{tab:detectorsData} separated by \textit{single spacing}. Other configurations can be added directly to the dictionary \texttt{detectors} in the module \texttt{gwfastGlobals}. Alternatively, one can pass a custom configuration with the option \texttt{netfile};
    \item[\texttt{psds}] \textcolor{brightgreen}{\texttt{list}} of \codekeyword{string}, default is \texttt{[}\codestring{ET-0000A-18.txt}\texttt{]}; the paths to PSDs of each detector in the network inside the folder \texttt{psds/}, separated by \textit{single spacing};
    
    \item[\texttt{netfile}] alternative to the use of \texttt{net} and \texttt{psds} to configure the detector network; a dictionary containing the configuration can be saved in \texttt{.json} format and passed as input. It is possible to save a network configuration as:
\begin{lstlisting}[caption={How to save a detector network configuration for \codename{}.},label={lst:save_det_conf}]
from gwfastUtils import save_detectors

my_network = {'my_detector_1': { 'lat': ..., 'long': ... , 'xax': ... ,
                                 'shape': ... , 'psd_path': 'path/to/psd'},
              'my_detector_2': {...}
             }
             
save_detectors('network_file_name.json', my_network)
\end{lstlisting}
    then, send run with \texttt{\textcolor{brightgreen}{--}netfile=network\_file\_name.json};
    \item[\texttt{mpi}] \codekeyword{int}, default is \textcolor{brightgreen}{\texttt{1}}; if \textcolor{brightgreen}{\texttt{0}}, the code parallelizes using \texttt{multiprocessing}, if \textcolor{brightgreen}{\texttt{1}}, it parallelizes using \texttt{MPI}, suitable for clusters. In this case, the function should be called accordingly, e.g.
\begin{lstlisting}[caption={How to parallelize a run using \texttt{MPI}.},label={lst:use_MPI}, language=bash, literate={-}{{{\color{brightgreen}-}}}1
{4}{{{{4}}}}1
{>}{{{\color{brightgreen}>}}}1]
> mpirun -n 4 python calculate_forecasts_from_catalog.py ... --mpi=1 --npools=4 
\end{lstlisting}
    \item[\texttt{duty\_factor}] \codekeyword{float} $\in [0,1]$, default is \textcolor{brightgreen}{\texttt{1.}}; duty factor of the detectors. This is applied separately to each detector in the network (and to each component separately in the case of a triangular configuration);
    \item[\texttt{params\_fix}] \textcolor{brightgreen}{\texttt{list}} of \codekeyword{string}, default is \texttt{[ ]}; parameters to fix to the fiducial values, i.e. to eliminate from the FIM;
    \item[\texttt{rot}] \codekeyword{int}, default is \textcolor{brightgreen}{\texttt{1}}; if \textcolor{brightgreen}{\texttt{0}} the effect of the rotation of the Earth is \textit{not} included in the analysis, if \textcolor{brightgreen}{\texttt{1}} it is included;
    \item[\texttt{lalargs}] \textcolor{brightgreen}{\texttt{list}} of \codekeyword{string}, default is \texttt{[ ]}; specifications of the waveform when using \texttt{LAL} interface. This has to contain \codestring{HM} if the waveform includes the contribution of higher--order modes, \codestring{tidal} if it contains tidal effects, \codestring{precessing} if it includes precessing spins, and \codestring{eccentric} if it includes eccentricity;
    \item[\texttt{return\_all}] \codekeyword{int}, default is \textcolor{brightgreen}{\texttt{1}}; if \textcolor{brightgreen}{\texttt{1}}, in case a network of detectors is used, the SNRs and Fishher matrices of the individual detector are also stored;
    \item[\texttt{seeds}]\textcolor{brightgreen}{\texttt{list}} of \codekeyword{int}, default is \texttt{[ ]}; list of seeds to set for the duty factors in individual detectors, to make the results easily reproducible.
\end{description}
To show the performance gain using vectorization, we report in \autoref{fig:Speed_test} the ratio $t/(N t_1)$ among the time $t$ needed to compute SNRs and Fisher matrices on $N$ events at the same time on the same CPU, and $N$ times the time $t_1$ needed to compute the same quantities for 1 event (which is the time needed using a \codekeyword{for} loop). From the left panel, referring to the SNRs, the impressive gain brought by vectorization is apparent, with an amount of time need for the computation that stays basically constant while enlarging the batch size, thus effectively being $N$ times faster than a loop--based computation. Quantifying the advantage from vectorization when computing Fisher matrices is instead more subtle: as it is apparent from the right panel of \autoref{fig:Speed_test}, for $N\gtrsim10$ the behaviour has a dependence on the characteristics of the machine used to run the code. Differently from SNRs, Fisher matrices need much more memory to be allocated during the computation, especially when vectorizing on $N$ events, and the operations can become much slower on these large arrays, eventually leading to a loss in terms of speed as compared to a serial computation, which instead handles smaller arrays. In any case, there is always an ‘optimal' batch size, depending on the machine's characteristics, such that the gain in terms of speed thanks to vectorization can be as large as a factor of $\sim 5$. 

\section{Comparison with real events}\label{sec:comparison_realEv}
\begin{figure}
    \centering
    \includegraphics[width=.92\textwidth]{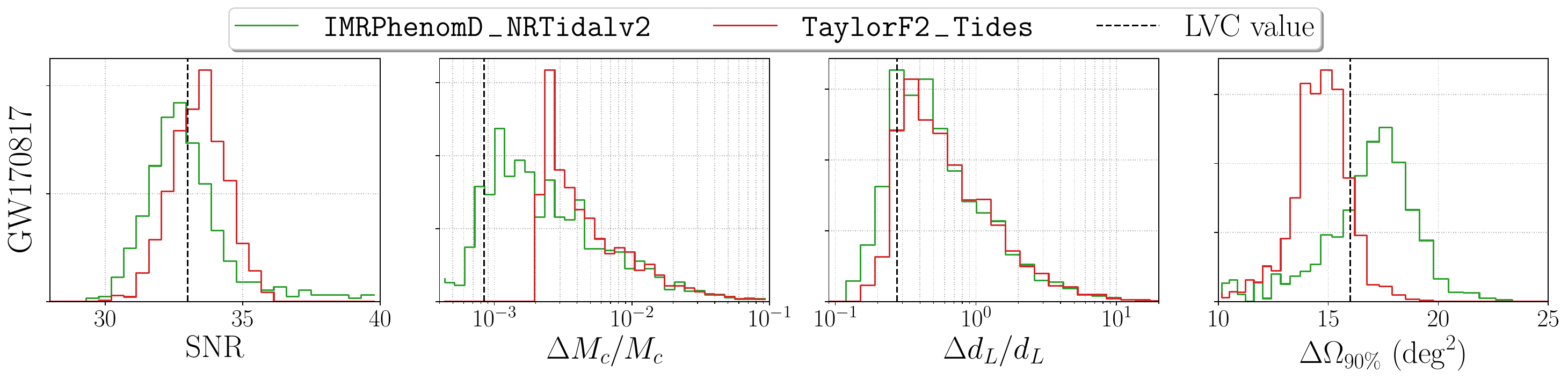}\\
    \includegraphics[width=.92\textwidth]{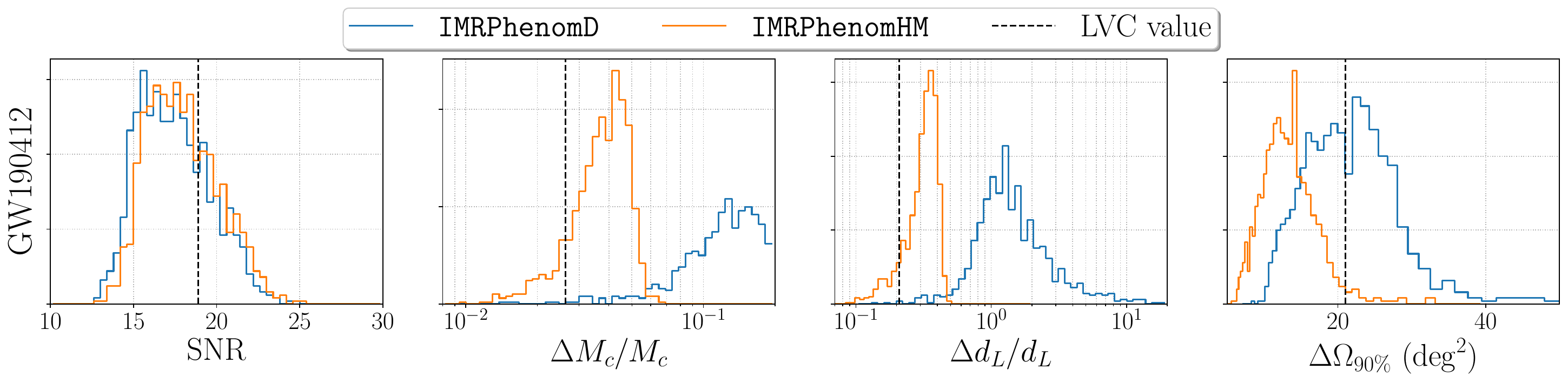}\\
    \includegraphics[width=.92\textwidth]{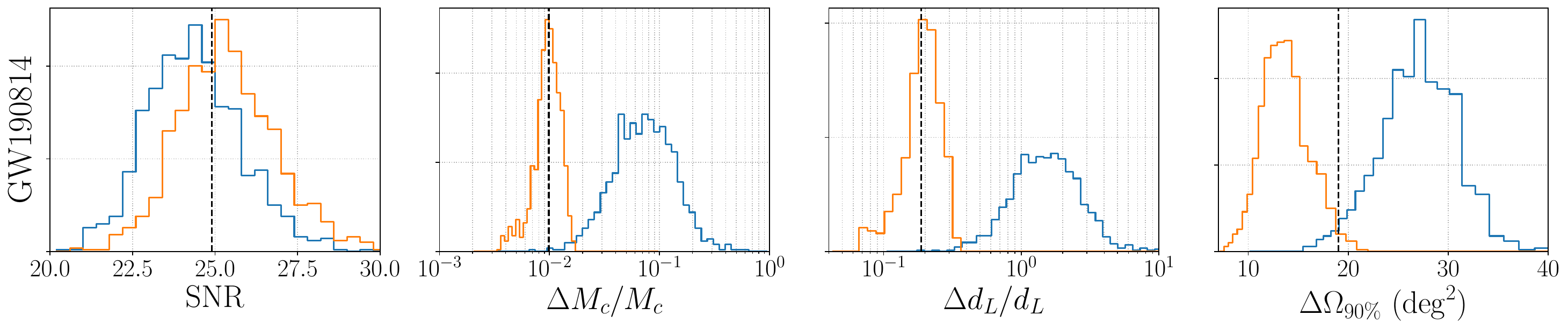}\\
    \includegraphics[width=.92\textwidth]{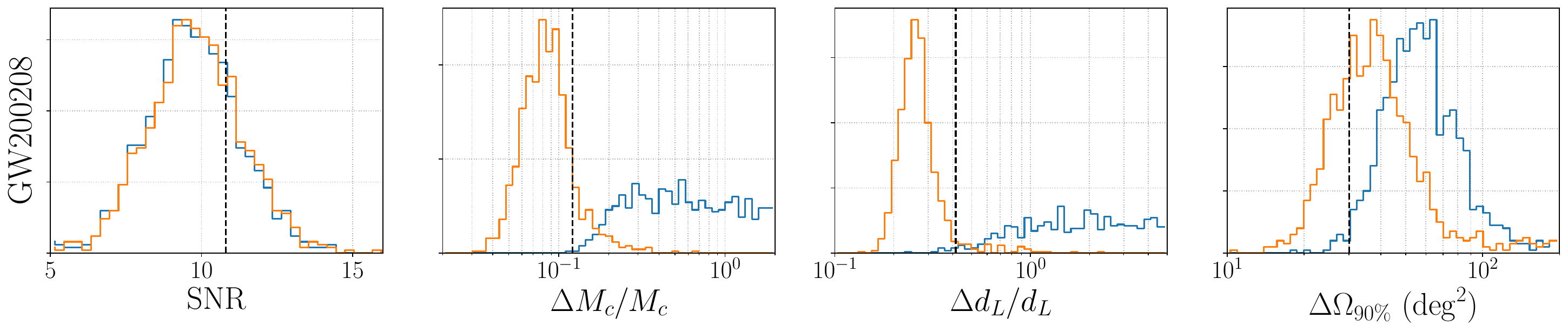}\\
    \includegraphics[width=.93\textwidth]{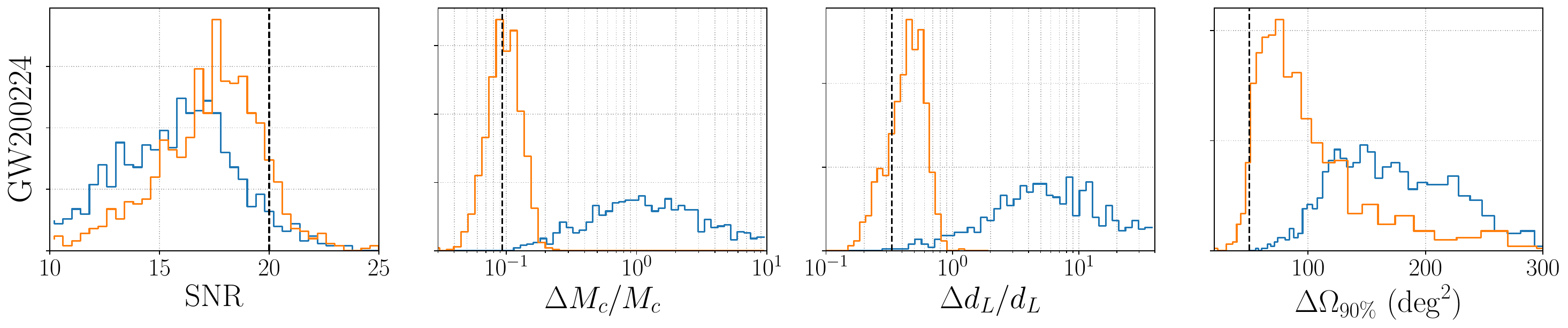}\\
    \includegraphics[width=.92\textwidth]{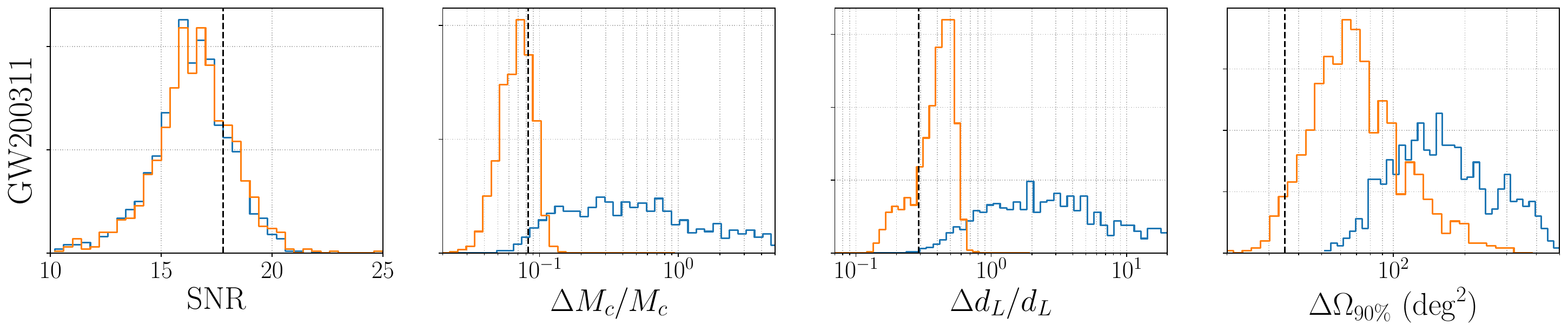}\\
    \caption{Comparison of the results obtained using \codename{} for some selected events from GWTC--2 and GWTC--3: GW170817, GW190412, GW190814, GW200208\_130117, GW200224\_222234 and GW200311\_115853. Each row contains the histogram of the SNR, $90\%$ relative credible intervals on the source--frame chirp mass and luminosity distance, and 90\%--credible sky area for 1000 samples of the posterior distributions of the events for all the parameters. Lines of different colours refer to different waveform models and the vertical dashed lines denote the errors inferred by LVC with a full Bayesian parameter estimation \citep{LIGOScientific:2020ibl,LIGOScientific:2021djp}.}
    \label{fig:comparison_real_ev}
\end{figure}

As an illustration of the reliability of \codename{}, we checked how its predictions compare to the SNRs and measurement errors associated to real events. This is a stringent test since, for currently available events, many assumptions on which the FIM approximation is based are of course not valid, in particular the Gaussianity of the noise and the limit of high SNR. Here we show that we can still reproduce correctly the order of magnitude of SNRs and relative errors, taking into account the broad measurement errors of the real events.
We selected a subset of the GW events in GWTC--1, GWTC--2 and GWTC--3, with ${\rm SNR} \geq 10$ and sky localization $\Delta\Omega_{90\%}\leq\SI{50}{\square\degfull}$.\footnote{In general, poor sky localization regions are associated to highly non--Gaussian posteriors, at least in the angular variables. Note that the requirement of having a large SNR does not necessarily imply a good localization, for which triangulation is essential: for example, the first event detected by LVC -- GW150914 -- has an SNR of $\sim 24$ but a rather large localisation region of about \SI{182}{\square\degfull}, having been observed only by the two LIGO detectors.} 
For each selected event we extracted 1000 samples from the full set of posterior samples\footnote{All the data is available on GWOSC, \url{https://www.gw-openscience.org}.\label{footnote:GWOSClink}} for each parameter.  We used ASDs obtained averaging the strain of each detector in a window of 1024 seconds around each event,\textsuperscript{\ref{footnote:GWOSClink}} using Welch's method, through the software \texttt{GWpy} \citep{Papergwpy}.

The results for GW170817 (BNS), GW190412, GW190814, GW200208\_130117, GW200224\_222234 and GW200311\_115853 (BBHs) are shown in \autoref{fig:comparison_real_ev}, where we compare the errors estimated from \codename{} on the set of 1000 samples drew for each event to the actual measurement errors obtained from the full Bayesian analysis (which can be read from Tab. III of \cite{LIGOScientific:2018mvr} for the events belonging to GWTC--1, Tab. VI of \cite{LIGOScientific:2020ibl} for those belonging to GWTC--2, and Tab. IV of \cite{LIGOScientific:2021djp} for those belonging to GWTC--3). In particular, we show the distribution of the SNRs, the 90\% relative credible intervals on the source--frame chirp mass, $M_c$, and luminosity distance\footnote{Computed, for ease of comparison, converting the $1\sigma$ errors obtained from the inversion of the FIM.} and the size of the 90\%--credible sky area, and report as a dashed line the LVK error estimate. For the BNS event GW170817 we performed the analysis with the waveform models \texttt{TaylorF2\_Tides} and \texttt{IMRPhenomD\_NRTidalv2}, both including tidal effects, while for the BBH events we used \texttt{IMRPhenomD} and \texttt{IMRPhenomHM}, which includes the contribution of higher modes, always taken into account in the parameter estimation of the chosen BBH signals \citep{LIGOScientific:2020stg, Abbott:2020khf, LIGOScientific:2021djp}. 

We find overall a very good agreement for the SNR distributions for all events, with both the waveform models used in each case, and we also observe that, as expected in the BBH case, \texttt{IMRPhenomHM} produces slightly higher SNRs than \texttt{IMRPhenomD}, especially for GW190814. This can be traced to the fact that the mass ratio of this system is large (the primary component has been estimated to have a source--frame mass of about \SI{23}{\Msun} and the secondary \SI{2.6}{\Msun}), resulting in a greater relevance of the sub--dominant modes, as compared to more symmetric binaries (see e.g. Fig. 2 of \cite{Puecher:2022sfm}).

Regarding GW170817, we find that the agreement of the fractional error on the chirp mass $M_c$ with the LVK estimate, despite the long inspiral, is better using the full inspiral--merger--ringdown model \texttt{IMRPhenomD\_NRTidalv2}, which was not included in the first analysis of the system, while the distributions of the fractional error on $d_L$ are similar, and the distribution of the sky localization $\Delta\Omega_{90\%}$ are compatible.

For the majority of the BBH systems, we find our estimations on the source--frame chirp mass and luminosity distance errors to be compatible with the values inferred by LVK when using the waveform model including higher--order harmonics, which were indeed included in the analysis. The only exception is the system GW200208\_130117, for which our estimations seem optimistic. This can be understood by the fact that the network SNR for this system (equal to 10.8) is the lowest among the ones considered: for such a value of the SNR, the FIM approach is not guaranteed to work. In any case, even in this case the sky localization is compatible and there is always a fraction of cases where also the fractional errors on chirp mass and distance are consistent.
As for the sky localisation, we find our estimations to include the LVK results for all the events when using \texttt{IMRPhenomHM}, always being on the same order of magnitude and without a clear trend towards higher or lower values for different events.

\section{Summary}\label{sec:summary}
In this article we presented \codename{}~\raisebox{-1pt}{\href{https://github.com/CosmoStatGW/gwfast}{\includegraphics[width=10pt]{GitHub-Mark.pdf}}}, a novel pure \texttt{Python} code for computing SNRs and Fisher information matrices for catalogs of GW events, in a fast, accurate and user--friendly way. In particular, \codename{}:
\begin{itemize}[label=--]
    \item implements a pure \texttt{Python} version of state--of--the--art Fourier domain full inspiral--merger--ringdown waveform models, suitable for both BBH, BNS and NSBH systems, namely \texttt{IMRPhenomD}, \texttt{IMRPhenomD\_NRTidalv2}, \texttt{IMRPhenomHM} and \texttt{IMRPhenomNSBH}. These are also separately available in the module \wfname{}~\raisebox{-1pt}{\href{https://github.com/CosmoStatGW/WF4Py}{\includegraphics[width=10pt]{GitHub-Mark.pdf}}} (which further includes the waveform model \texttt{IMRPhenomXAS}), and allow to exploit vectorization to speed up the computation and to employ automatic differentiation for computing derivatives. It is also possible to use all waveforms included in \texttt{LAL}, in which case derivatives are computed with finite differences techniques;
    \item accounts for the amplitude and phase modulation of the observed GW signal due to Earth's rotation, which is of fundamental importance at 3G detectors, whose sensitivity curve can extend down to \SI{2}{\hertz}, in particular for BNS systems, which can stay in the detection band for as long as $\order{\SI{1}{\day}}$;
    \item is developed to handle networks of detectors, both L--shaped and triangular, and includes 10 pre--defined locations as well as several sensitivity curves, for both current and planned ground--based detectors, which can also easily be extended;
    \item if waveforms in \texttt{Python} are used, computes derivatives using \emph{automatic differentiation}, through the \texttt{JAX} package, thus being extremely accurate and fast, and offers the possibility of computing derivatives with respect to many parameters analytically, to further speed up the computation;
    \item handles the inversion of the FIM using the \texttt{mpmath} library, thus avoiding limitations linked to numerical precision, and includes functions for easily manipulating both the Fisher and covariance matrices, e.g. adding priors or computing localisation regions, as well as tools to assess the reliability of the inversion;
    \item can compute SNRs and Fisher matrices for multiple events at a time \emph{on a single CPU} exploiting \texttt{Python} vectorization, and provides a module for parallelization over multiple CPUs, also suitable for clusters, thus being ideal for large catalogs of sources;
\end{itemize}
We assessed the reliability of \codename{} in computing accurately the signal derivatives, as well as the frequency integral, by comparing with an independent code written in \texttt{Wolfram Mathematica}, capable of computing analytical derivatives with respect to all parameters, obtaining excellent agreement. We further compared the predictions for the SNRs and measurement errors obtained using \codename{} on some of the loudest and best localised events detected during the second and third observing runs of the LVK collaboration with the actual results obtained from a full Bayesian parameter estimation, obtaining good agreement. 
\codename{} has been used to produce the results in the companion paper \citep{Iacovelli:2022bbs}, where we also discuss its comparison with other existing codes and results \citep{Borhanian:2020ypi,Pieroni:2022bbh,Harms:2022ymm}, showing their excellent agreement.
Due to its structure and to the use of automatic differentiation, \codename{} is also suitable for extensions of the FIM approximation \citep{Sellentin:2014zta, Vallisneri:2011ts, Wang:2022kia}.
We are confident that it will constitute a useful tool for assessing the scientific potential of third--generation GW detectors.  
\codename{} \raisebox{-1pt}{\href{https://github.com/CosmoStatGW/gwfast}{\includegraphics[width=10pt]{GitHub-Mark.pdf}}} is publicly available at \url{https://github.com/CosmoStatGW/gwfast}. This paper is associated to version v1.0.1 which is archived on Zenodo \citep{gwfast_zenodo}.
The library \wfname{} \raisebox{-1pt}{\href{https://github.com/CosmoStatGW/WF4Py}{\includegraphics[width=10pt]{GitHub-Mark.pdf}}} is available at \url{https://github.com/CosmoStatGW/WF4Py}. This paper is associated to version v1.0.0 which is archived on Zenodo \citep{WF4Py_zenodo}.

\begin{acknowledgments}
{\em Acknowledgements}.
Our research is supported by  the  Swiss National Science Foundation, grant 200020$\_$191957, and  by the SwissMap National Center for Competence in Research.
The research leading to these results has been conceived and developed within the ET Observational Science Board (OSB).
\end{acknowledgments}

\bibliography{myrefs}{}
\bibliographystyle{aasjournal}

\end{document}